\documentclass{aa}  

\usepackage{graphicx}
\usepackage{xcolor}
\usepackage{txfonts}
\newcommand{\Msun}{M_{\odot}}

\newcommand{\lcdm}{$\Lambda$cold dark matter}
\newcommand{\phox}{\textsc{Phox}}

\newcommand{\xspec}{\texttt{XSPEC}}

\newcommand{\usfr}{\Msun\,\mathrm{yr}^{-1}}
\newcommand{\ulum}{\mathrm{erg\, s^{-1}}}

\begin{document}

   \title{Decomposition of galactic X-ray emission with \phox{} }

   \subtitle{Contributions from hot gas and X-ray binaries}

   \author{S. Vladutescu-Zopp
          \inst{1}\fnmsep\thanks{\email{vladutescu@usm.lmu.de}}
          \and
          V. Biffi\inst{2} %\fnmsep\thanks{Just to show the usage of the elements in the author field}
          \and
          K. Dolag\inst{1}\fnmsep\inst{3}
          }
          
    \titlerunning{Galactic X-ray emission in PHOX}
    \authorrunning{Vladutescu-Zopp S., et al.}

   \institute{Universitäts-Sternwarte, Fakultät für Physik, Ludwig-Maximilians-Universität München, Scheinerstr.1, 81679 München, Germany 
             \and
             INAF, Osservatorio Astronomico di Trieste, via Tiepolo 11, I-34131, Trieste, Italy
             \and
             Max-Planck-Institut f\"ur Astrophysik, Karl-Schwarzschild-Straße 1, 85748 Garching bei M\"unchen, Germany
             }

   \date{Received 08/2022; accepted 09/11/2022}

% \abstract{}{}{}{}{} 
% 5 {} token are mandatory
 
  \abstract
   {X-ray observations of galaxies with high spatial resolution instruments such as \textit{Chandra}  have revealed that major contributions to their diffuse emission originate from X-ray-bright point sources in the galactic stellar field. It has been established that these point sources, called X-ray binaries, are accreting compact objects with stellar donors in a binary configuration. They are classified according to the predominant accretion process: wind-fed in the case of high-mass donors and Roche-lobe mass transfer in the case of low-mass donors. Observationally, it is challenging to reliably disentangle these two populations from each other because of their similar spectra. }
   {We provide a numerical framework with which spatially and spectrally accurate representations of X-ray binary populations can be studied from hydrodynamical cosmological simulations. We construct average spectra, accounting for a hot gas component, and verify the emergence of observed scaling relations  between galaxy-wide X-ray luminosity ($L_X$) and stellar mass ($M_*$) and between $L_X$ and the star-formation rate (SFR).
   }
   {Using simulated galaxy halos extracted from the $(48\,h^{-1} \mathrm{cMpc})^3$ volume of the Magneticum Pathfinder cosmological simulations at $z=0.07$, we generate mock spectra with the X-ray photon-simulator \phox{}. We extend the \phox{} code to account for the stellar component in the simulation and study the resulting contribution in composite galactic spectra.}
   {Well-known X-ray binary scaling relations with galactic SFR and $M_*$ emerge self-consistently, verifying our numerical approach. Average X-ray luminosity functions are perfectly reproduced up to the one-photon luminosity limit. Comparing our resulting $L_X-\mathrm{SFR}-M_*$ relation for X-ray binaries with recent observations of field galaxies in the Virgo galaxy cluster, we find significant overlap. Invoking a metallicity-dependent model for high-mass X-ray binaries yields an anticorrelation between mass-weighted stellar metallicity and SFR-normalized luminosity. The spatial distribution of high-mass X-ray binaries coincides with star-formation regions of simulated galaxies, while low-mass X-ray binaries follow the stellar mass surface density. X-ray binary emission is the dominant contribution in the hard X-ray band (2-10 keV) in the absence of an actively accreting central super-massive black hole, and it provides a $\sim 50\%$ contribution in the soft X-ray band (0.5-2 keV), rivaling the hot gas component.}
   {We conclude that our modeling remains consistent with observations despite the uncertainties connected to our approach. The predictive power and easily extendable framework hold great value for future investigations of galactic X-ray spectra.}

   \keywords{X-ray: galaxies - X-ray: binaries - X-ray: ISM - methods: numerical}

   \maketitle
%
%-------------------------------------------------------------------

\section{Introduction}

   X-ray properties of local normal galaxies have been extensively studied in the past, where ``normal'' describes galaxies without a luminous active galactic nucleus (AGN). One of the most challenging tasks has been the identification of the physical processes that produce the observed X-ray emission from those galaxies. It is a well-established fact that most of their X-ray power-output originates from diffuse hot gas in the interstellar medium (ISM) and accretion-powered point sources called X-ray binaries \citep[XRBs; see the reviews][]{Fabbiano1989,Fabbiano+2006,Fabbiano2019}, excluding the emission from actively accreting super-massive black holes (SMBHs).
   
   Numerous studies of the hot diffuse ISM in star-forming normal galaxies have revealed a tight linear relation between the total X-ray luminosity, $L_X$, and the star-formation rate \citep[SFR;][]{Strickland+2000,Ranalli+2003,Tyler+2004,Mineo+2012b,Li+Wang2013b,Kouroumpatzakis+2020} due to feedback from young stellar objects and energy injection from supernovae (SNe), which heat up the ISM. Recent studies suggest that the ISM emission normalization in X-rays declines with increasing total gas-phase metallicities, possibly due to the effect of metal absorption and changes in stellar feedback within the ISM \citep{Garofali+2020,Lehmer+2022}.
   
   The ISM of normal elliptical galaxies (EGs) depends on their dynamical state \citep{Kim+Fabbiano2015,Babyk+2018} and strongly correlates with the ISM temperature \citep{Boroson+2011,Kim+Fabbiano2015} and dynamical mass \citep{Kim+Fabbiano2013,Forbes+2017,Babyk+2018}. The ISM of core EGs, where ``core'' denotes slow rotating systems with cored central surface brightness and an overall old stellar population, appears to follow gas X-ray luminosity scaling relations consistent with virialized systems, similar to brightest cluster galaxies (BCGs) and groups. Other types of gas-poor EGs behave similarly to disk galaxies in these relations: they do not show significant correlations between hot gas temperature and the X-ray luminosity of hot gas, suggesting secondary effects such as rotation, flattening, and SN feedback dissipating the ISM \citep[see, e.g.,][]{Fabbiano2019}.
   
   Separate scaling relations are found for collective XRB X-ray luminosity depending on the type of XRB: low-mass X-ray binaries (LMXBs) are accreting compact objects (COs) where the low-mass companion undergoes Roche-lobe overflow at the end of its life. Thus, LMXBs are generally correlated with the integrated stellar light ($L_k$) and, by extension, with the total stellar mass of the host galaxies because of the long evolutionary timescales of the low-mass ($<1\,\Msun$) donor star in the binary system \citep{Gilfanov2004,Boroson+2011,Zhang+2012,Lehmer+2019}. This relation seems to be enhanced by the globular cluster (GC) specific frequency, $S_N=N_{\mathrm{GC}}10^{0.4(M_V^T+15)}$, where gravitationally dissolving GCs seed LMXBs into the galactic field, which join the LMXBs formed in situ  \citep{Irwin2005,Zhang+2011,Boroson+2011,Zhang+2012,Lehmer+2020}.
   In high-mass X-ray binaries (HMXBs) the donor star is a massive O/B star $(>8\,\Msun)$ that fuels accretion onto the CO with intense stellar winds. Due to the short lifetimes of such massive stars $(<100 \mathrm{Myr})$, HMXBs provide an independent tracer of the galactic star-formation history (SFH). It was found that the combined X-ray luminosity of HMXBs relates linearly to the SFR \citep{Grimm+2003,Shty+2005b,Mineo+2012a,Lehmer+2010,Lehmer+2016,Lehmer+2019}.
   Broadband X-ray luminosity functions (XLFs) of XRBs follow distinct power-law (PL) or broken-power-law (BPL) distributions in the local Universe \citep[see, e.g.,][]{Grimm+2003,Gilfanov2004}. The exact shape of XRB luminosity functions in normal galaxies is a function of stellar age and evolves on timescales consistent with the stellar evolution timescales of the XRB donor, transitioning from HMXB dominated to LMXB dominated at a stellar population age of $\gtrsim 100\,\mathrm{Myr}$ \citep{Lehmer+2017,Gilbertson+2022}.
   
   By combining the scaling relations of both XRB types, we expect a distinct relation between SFR-normalized $L_X$ and the specific star-formation rate (sSFR = SFR/$M_*$) of the form $L_X/\mathrm{SFR} = \alpha\, \mathrm{sSFR}^{-1}+\beta$, where $\alpha$ describes the contribution from LMXBs and $\beta$ the contribution from HMXBs. This relation is especially useful in galaxies where XRB candidates cannot be resolved: it shows which of the two XRB types is the dominant contributor to the galactic $L_X$. In previous works it has been shown that this relation generally holds for local galaxies \citep{Lehmer+2010,Lehmer+2016,Lehmer+2019}, with modifications for high redshift galaxies \citep{Lehmer+2016}. The turnover from LMXB to HMXB domination occurs at $\log\mathrm{sSFR}\sim-10.5$. Recently, \citet{Soria+2022} performed \textit{Chandra} observations of normal galaxies in the Virgo cluster that were marginally compatible with scaling relations for local galaxies, indicating non-negligible environmental effects on this relation. A similar dependence was found by \citet{Inoue+2021} in the form of a fundamental plane: $L_X \sim \mathrm{SFR}+\alpha\,M_*$.
   
   Numerous studies on generalizing locally derived XRB scaling relations for the high redshift Universe have used computationally expensive population synthesis codes for XRB evolution \citep{Fragos+2013a,Fragos+2013b,Madau+Fragos2017,Wiktorowicz+2017}. Taking into account the metallicity evolution and SFH of the universe constrained with the help of cosmological simulations (\textsc{Millenium II} in the case of \citealt{Fragos+2013b}) suggests a more complex evolution of local XRB scaling relations, with redshift and HMXB emissivity dominating above $z\sim2$, where the cosmic SFR density peaks. This is supported by constraints from the VANDELS survey on HMXB emission in high redshift galaxies \citep{Saxena+2021}. {The VANDELS project is an ESO funded spectroscopic survey within the \textit{Chandra Deep Field South} targeting high-redshift galaxies.} Additional evidence for the metallicity dependence of HMXB emissivity comes from a recent study by \citet{Lehmer+2021}, who used the gas-phase oxygen fraction of their galactic sample to further refine the $L_X$-SFR relation. They demonstrate that galaxies with lower metal content consistently show an increased number of luminous HMXBs and a higher total luminosity. Similarly, earlier studies from \citet{Brorby+2016} and \citet{Fornasini+2020} find anticorrelated metallicity enhancements in the $L_X$-SFR relation, which is consistent with findings by \citet{Lehmer+2021}. Additionally, data from the eROSITA Final Equatorial Depth Survey (eFEDS) show an elevated total X-ray luminosity for low-metallicity dwarf galaxies with high sSFRs \citep{Vulic+2021}.
   This paper introduces a numerical modeling to study the composition of galactic X-ray emission from a theoretical point of view, using state-of-the-art hydrodynamical cosmological simulations from the Magneticum set. Instead of applying computationally expensive population synthesis codes, we associated XRB populations with the stellar component within the cosmological simulation by making use of observationally derived XRB luminosity functions. We use observationally derived local scaling relations to constrain seeding parameters and describe the numerical setup used to extend the capabilities of the virtual X-ray photon simulator \phox{}. \phox{} has been successfully used in conjunction with Magneticum to study galaxy cluster $L_X$-temperature relations as well as AGN luminosity functions. Additionally, it was used to predict the contamination of cluster X-ray emission by AGN for the eROSITA mission \citep[see][]{Biffi+2012,Biffi+2013,Biffi+2018b}. The inclusion of an XRB component enables the study of galactic X-ray spectra from a theoretical standpoint. We validate our approach by retrieving well-known XRB and gas scaling relations of galaxies while maintaining a low computational cost. 
   
   The paper is structured as follows:
   In Sect. \ref{sec:Magneticum} we describe the state-of-the-art hydrodynamical cosmological simulation suite Magneticum, on which we base our analysis. In Sect. \ref{sec:SSPs} we describe the methodology we used to constrain the XRB population within the stellar component of cosmological simulations and a novel approach to estimate linear SFRs for simple stellar populations (SSPs). In Sect. \ref{sec:Phox} we describe the process for generating photons from X-ray-emitting sources in the simulations using the virtual photon simulator \phox{}. In Sect. \ref{sec:Results} we show our results regarding the reconstruction of known XRB and hot gas scaling relations as well as a proper spatial  distribution of the two XRB types in the galactic field. In Sect. \ref{sec:Counts} we compare average count ratios of different contributors to galactic X-ray emission and their relative contribution to the total X-ray emission. In Sect. \ref{sec:Discussion} we put our findings into context with our methodology and constructed data set by highlighting caveats and uncertainties. Finally, we summarize our findings in Sect. \ref{sec:Conclusions}.  

\section{Cosmological hydrodynamical simulation}
   \label{sec:Magneticum}

   Magneticum Pathfinder simulations\footnote{Project web page: \texttt{www.magneticum.org}} are a series of state-of-the-art hydrodynamical cosmological simulations
   that explore varying ranges in particle number, volume, and resolution. They are based on an improved version of the
   N-body code \textsc{Gadget 3}, which is an updated version of the code \textsc{Gadget 2} (\cite{Springel2005})
   including a Lagrangian method for solving smoothed particle hydrodynamics (SPH). The code introduces several improvements regarding the SPH implementation by including a
   treatment of viscosity and artificial conduction \citep{Dolag+2005,Beck+2016}. Additional physical processes describing the evolution of the baryonic component have been implemented. They encompass radiative gas cooling as described in \citet{Wiersma+2009}, heating from a uniform time-dependent UV background \citep{Haardt+2001} and a sub-resolution model for star-formation with mass-loading rate proportional to SFR and resulting wind-velocities of $v_w = 350\,km\,s^{-1}$\citep{SH2003}. A treatment for chemical enrichment of the gaseous component through stellar evolution has been implemented following the prescription of \citet{LT+2004, LT+2007}. Furthermore, a prescription of SMBH growth and gas accretion, powering energy feedback for AGN was implemented following \citet{Springel2005} and \citet{DiMatteo+2005} with modifications following \citet{Fabjan+2010}. Previous studies using the Magneticum simulations are consistent with observed kinematic and morphological properties of galaxies \citep{Teklu+2015,Teklu+2017,Remus+2017a,Schulze+2018,Schulze+2020}, chemical properties of galaxies and clusters \citep{Dolag+2017} as well as statistical properties of AGN \citep{Hirschmann+2014, Steinborn+2016,Biffi+2018b}.
   In this paper we make use of a high resolution run of Magneticum called Box4/uhr, which is a $(48\,h^{-1}\,\mathrm{cMpc})^3$ comoving volume with a mass resolution of $m_{\mathrm{DM}} = 3.6\times10^7\,\Msun$ and $m_{\mathrm{gas}} = 7.3\times10^6\,\Msun$ for dark matter and gas, respectively, and a total of $576^3$ particles.
   Initial conditions for the simulations are generated using standard \lcdm{} cosmology using results from
   the {\textit{Wilkinson} Microwave Anisotropy Probe} \citep[WMAP7;][]{WMAP7}, with Hubble parameter $h=0.704$,
   matter density $\Omega_M = 0.272$, dark energy density $\Omega_{\Lambda} = 0.728$, baryon density $\Omega_b=0.0451$
   and normalization of the fluctuation amplitude at 8 Mpc $\sigma_8 = 0.809$.
   
\section{Modeling of XRB populations within SSPs} \label{sec:SSPs}
   
   In this section we describe the process of modeling the population size of
   LMXBs and HMXBs, given a stellar resolution element in the cosmological simulation. Stellar elements describe a collection of stars as a SSP that shares the same time of birth and initial metallicity, distributed according to a common initial mass function (IMF). The main difficulty lies
   in connecting the XLFs, which are derived from observed galactic
   properties, with SSP properties traced by the simulation. In order to preserve the 
   generality of our approach, we do not a priori assume a galactic environment of the SSP in
   the simulation but rather investigate scaling relations and arising XLFs by processing each SSP individually. 
   
   \subsection{Theoretical estimation of the LMXB population size}
   The modeling of LMXB is based on the fact that they cannot be associated with previous
   star formation events in a galaxy, which is due to the long lifetime of their low-mass
   stellar companions. Thus, LMXBs are more typically found in old galactic stellar regions,
   which connects the LMXB population to the total stellar mass of their host galaxy 
   \citep{Gilfanov2004, Zhang+2012, Lehmer+2017, Lehmer+2019}. By requiring that a SSP must
   be older than 1 Gyr for it to be eligible to host a LMXB population, we can simply obtain
   the LMXB population size by re-normalizing the LMXB XLF to the mass of the SSP. In particular,
   we use the BPL XLF from \cite{Zhang+2012} throughout this paper, which
   has the form   \begin{equation} \label{eq:xlf_LMXB}
      \dfrac{dN_{\mathrm{LMXB}}}{dL} = A_{\mathrm{LMXB}}\dfrac{M_*}{10^{11}\Msun}\begin{cases}
      L^{-\alpha_1}, &L < L_{b,1}\\
      L_{b,1}^{(\alpha_2-\alpha_1)}L^{-\alpha_2}, &L_{b,1} < L < L_{b,2}\\
      L_{b,2}^{(\alpha_3-\alpha_2)}L^{-\alpha_3}, &L_{b,2} < L < L_{cut}\\
      0\,, & L \geq L_{cut}
      \end{cases}\, ,
   \end{equation}
   where $L_{b,1}=0.546$ and $L_{b,2}=5.99$ are break luminosities, $L_{cut}=500$ is the cutoff luminosity,
   $A_{\mathrm{LMXB}}=54.48$ is the normalization, $M_*$ is the total stellar mass of a galaxy, and $\alpha_1=1.02$,
   $\alpha_2=2.06$, $\alpha_3=3.63$ are the respective PL slopes. Luminosities are given in units of $10^{38}\, \ulum$.
   In fact, LMXBs appear to not only be dependent on the stellar mass of the host galaxy,
   but also be connected to the GC specific frequency ($S_N$), which enhances the XLF normalization by accounting for dynamically formed LMXBs in GCs \citep{Irwin2005, Boroson+2011, Zhang+2012, Lehmer+2020}.
   In \citet{Lehmer+2020} they were successful in estimating the field LMXB contribution from different formation channels while earlier studies like \citet{Zhang+2012} combine these channels into a single XLF.
   Since it is not feasible to account for GCs in the context of a cosmological
   simulation, because of mass resolution limits, integration of Eq. \eqref{eq:xlf_LMXB} is
   sufficient to describe the LMXB population in a SSP as
   \begin{equation} \label{eq:tNLMXB}
       \tilde{N}_{\mathrm{LMXB}} = \int_{L_{\mathrm{min}}}^{L_{cut}} \dfrac{dN_{\mathrm{LMXB}}}{dL}(M_{\mathrm{SSP}})\, dL\, ,
   \end{equation}
   where we use the current SSP mass $M_{\mathrm{SSP}}$ in solar masses instead of $M_*$. The total stellar mass of a galaxy would then be recovered by the sum over each single SSPs within its boundary.
   
   \subsection{Theoretical estimation of the HMXB population size}
   
   Due to the short lifetime of their massive companions, HMXBs are typically associated with
   young stellar regions of a galaxy, which implies a connection to the SFR. It was shown that
   HMXBs are a robust independent tracer for recent star formation activity in their host galaxy
   \citep{Grimm+2003, Mineo+2012a}. In particular, the XLF normalization of HMXBs depends
   linearly on the SFR. A standard HMXB XLF that we use throughout the paper is given in
   \cite{Mineo+2012a} as
   \begin{equation} \label{eq:xlf_HMXB}
       \dfrac{dN_{\mathrm{HMXB}}}{dL} = A_{\mathrm{HMXB}}\dfrac{\mathrm{SFR}}{\usfr}\,\begin{cases} L^{-\gamma}\, , &L<L_{cut}\\
       0\, , &L\geq L_{cut}\\
       \end{cases}\, ,
   \end{equation}
   where $A_{\mathrm{HMXB}}$ is the normalization constant, SFR of a galaxy is given in units of $\usfr$, 
   $\gamma=1.59$ is the PL slope, and $L_{cut}=10^3$ is the cutoff luminosity. 
   The luminosity $L$ is given in units of $10^{38}\,\ulum$.
   One can obtain the number of individual HMXBs in a galaxy with
   \begin{equation}
       \label{eq:NHMXB}
       N_{\mathrm{HMXB}} = \int_{L_\mathrm{min}}^{L_{cut}} \dfrac{dN_{\mathrm{HMXB}}}{dL} dL\, .
   \end{equation}
   
   The main difficulty in transferring the SFR scaling property to SSPs in a cosmological 
   simulation lies in determining the SFH of these SSPs. Since SSPs are created in an
   instantaneous star-formation event in the simulation \citep{Springel+2005}, their SFH resembles a $\delta$-function
   at their time of creation \citep{LT+2007}. To circumvent this issue, we can adapt
   the procedure in \cite{Mineo+2012a} by obtaining the fraction $f_X$ of COs
   becoming a XRB, which is derived from the birth rate of massive stars ($>8\,\Msun$) in a star-formation
   event.
   Instead of using the birthrate $\dot{N}_{\mathrm{CO}}$ of COs following a star-formation event,
   which is directly dependent on the IMF and SFR, we can equally use the type II supernova
    (SNII) rate $R_{\mathrm{SNII}}$, which hides the SFR dependence in the stellar
   lifetime function. Following \cite{LT+2007}, we can express the SNII rate as
   \begin{equation} \label{eq:RSNII}
       R_{\mathrm{SNII}}(\tau) = \Phi(m(\tau))\times \dfrac{dm(\tau)}{d\tau}\, ,
   \end{equation}
   where $\Phi(m)$ is a mass-normalized IMF, which gives the number of stars given
   the mass $m$, and $m(\tau)$ is the inverse stellar lifetime function of \cite{PM1993} (PM93), which
   gives the mass of stars dying at an age of $\tau$. The SNII rate $R_{\mathrm{SNII}}$ is given in units of $\mathrm{Gyr}^{-1}$ for a SSP of mass $1\,\Msun$. The PM93 lifetime function has the form
   \begin{equation} \label{eq:PMlt}
       \tau(m) = \begin{cases}
       10^{\left(1.338-\sqrt{1.79-0.2232\cdot(7.764 - \log m})\right)/0.1116 - 9}\, , & m \leq 6.6\,\Msun \\
       1.2\cdot m^{-1.85} + 0.003\, , & m > 6.6\,\Msun
       \end{cases} ,
   \end{equation}
   with mass $m$ of the star given in $\Msun$ and $\tau$ in units of Gyr.
   We employed the modified Chabrier IMF \citep{Chabrier2003} used in the simulation, which has the form
   \begin{equation}
       \Phi(m) \propto \begin{cases}
       m^{-2.3}\, , & 1.0\, \Msun < m < 100\, \Msun\\
       m^{-1.8}\, , & 0.5\, \Msun < m < 1.0\, \Msun\\
       m^{-1.2}\, , & 0.1\, \Msun < m < 0.5\, \Msun\\
       \end{cases}\, .
   \end{equation}
   Using the SNII rate, we followed the calculations in \cite{Mineo+2012a} to obtain an estimate of the HMXB population size.
   
       First we determined the product of the X-ray-bright fraction, $f_X$, and the average bright-phase duration, $\bar{\tau}_X$, assuming an XLF
       \begin{equation} \label{eq:fX}
           f_X\bar{\tau}_X \sim \dfrac{N_{\mathrm{HMXB}}(>\!L_{\mathrm{min}})}{\dot{N}(>\!8\,\Msun)}\, ,
       \end{equation}
       where $\dot{N}(>\!8\,\Msun)$ is the birthrate of massive stars. The birthrate can be obtained from the IMF to the first order using
       \begin{equation} \label{eq:dotN}
           \dot{N}(>\!8\,\Msun) = \dfrac{\int_8^{M_{\mathrm{u}}} \Phi(m)\,dm}{\int_{M_{\mathrm{l}}}^{M_{\mathrm{u}}} \Phi(m)\,m\,dm}
           \times \mathrm{SFR}\, ,
       \end{equation}
       with integration limits defined by the IMF. \citet{Mineo+2012a} determined instead the term in Eq. \eqref{eq:fX} as $f_X \sim 0.18\frac{0.1\,\mathrm{Myr}}{\bar{\tau}}$, based on binary evolution
       calculations for the most common type of Be HMXB and assuming $\bar{\tau}_X\sim 0.1\,\mathrm{Myr}$, therefore making the factor explicitly dependent on the bright phase duration.
       In our case, by computing directly the product of $f_X$ and $\bar{\tau}$ as in Eq. \eqref{eq:fX}, we can conveniently bypass uncertainties 
       connected to assumptions for the bright phase duration.
       Given the proportionality to SFR in Eq.~\eqref{eq:dotN} and of  $N_{\mathrm{HMXB}}$ (see Eq. \eqref{eq:NHMXB}), then the dependence of $f_X\bar{\tau}_X$ on SFR disappears.
       
       Then, from Eq. \eqref{eq:fX}, we can replace the birthrate,
       $\dot{N}$, with the SNII rate, $R_{\mathrm{SNII}}$, given the age of the SSP, and multiply by the
       current SSP mass to get the expected HMXB population size:
       \begin{equation} \label{eq:tNHMXB}
           \tilde{N}_{\mathrm{HMXB}} = \begin{cases}
           M_{\mathrm{SSP}}\times
           R_{\mathrm{SNII}}(t_{\mathrm{SSP}})f_X\bar{\tau}_X\, , & t_{\mathrm{SSP}} \leq \tau(8\,\Msun)\\
           0\, , & t_{\mathrm{SSP}} > \tau(8\,\Msun)
           \end{cases} .
       \end{equation}
       Additionally, we require $t_{\mathrm{SSP}}$ to be smaller than the lifetime of
       stars with mass $8\,\Msun$, in line with mass limits for HMXBs in the literature \citep[see, e.g.,][]{Lewin+2006}, which, according to Eq. \eqref{eq:PMlt}, gives
       $\approx30\,\mathrm{Myr}$. 
       Therefore, in the simulation, a stellar particle 
       older than 30 Myr should no longer represent a SSP containing stars with masses greater than $8\,M_{\odot}$. The same procedure can be repeated for different IMFs and XLF models.
   
In order to study the metallicity dependence of HMXBs, we also included the  model proposed by \citet[][]{Lehmer+2021} (L21):
   \begin{equation}\label{eq:xlf_hzb}
       \dfrac{dN_{\mathrm{HZB}}}{dL} = A_{\mathrm{HZB}}\,\mathrm{SFR}\,\exp[-L/L_c(Z)]
       \begin{cases}
       L^{-\gamma_1}\, , &L<L_{b}\\
       L_b^{\gamma_2(z)-\gamma_1}L^{\gamma_2(Z)}\, , &L\geq L_{b}\\
       \end{cases}\, ,
   \end{equation}
   with
   \begin{equation}
       \gamma_2(Z) = \gamma_{2,\odot} + \dfrac{d\gamma_2}{d\log Z}[12+\log(\mathrm{O/H})-8.69]\, 
   \end{equation}
   and
   \begin{equation} \label{eq:LcZ}
       \log L_c(Z) = \log L_{c,\odot} + \dfrac{d\log L_c}{d\log Z}[12+\log(\mathrm{O/H})-8.69]\, .
   \end{equation}
   Both $\gamma_{2,\odot} = 1.16$ and $\log L_{c,\odot} = 1.98$ are reference values at solar metallicity. 
   The other parameter values are the normalization $A_{\mathrm{HZB}} = 1.29$, break luminosity
   $\log L_b = 0.54$, slope $\gamma_1 = 1.74$ and first-order metallicity corrections
   $\frac{d\gamma_2}{d\log Z} = 1.34$ and $\frac{d\log L_c}{d\log Z} = 0.6$.
   Luminosities are given in units of $10^{38}\,\ulum$ in Eq. \eqref{eq:xlf_hzb}.
   We labeled the L21 model ``HZB'' to indicate the metallicity dependence, $Z$.
   
\section{Synthetic X-ray emission} \label{sec:Phox}
    
   In this section we outline the procedure to obtain synthetic X-ray emission
   of the hot ISM, AGN, and XRBs from the simulations, where we focus on
   the XRB emission model.
   
   \subsection{PHOX X-ray photon simulator}
   In this work we generated synthetic X-ray observations of the hot gas, actively accreting SMBHs,
   and the stellar component in the simulations using the \phox{} code
   \citep[see][for further details]{Biffi+2012,Biffi+2013}). In summary, the X-ray photon simulator \phox{}
   consists of three individual modules.
   
   \textbf{Unit 1}: For every emitting source in the simulation, an idealized emission is computed
       from a model spectrum, which is statistically sampled by a discrete number of photons. The generated
       photon data are stored such that they reflect the position and velocity information of the emitting sources
       in the simulation.

    \textbf{Unit 2}: The photon data are projected along a direction through the simulation
       and energies are Doppler shifted according to the line-of-sight velocity of the corresponding
       emitting source. Additionally, a spatial selection can be considered.
       
    \textbf{Unit 3}: The produced photon list is convolved with the specific response of a chosen
       X-ray telescope, with realistic observing time and detector area.
   %\end{itemize}
   
   The \phox{} code has been applied to simulations of galaxy clusters to study the properties of the
   hot diffuse {intra-cluster medium} \citep{Biffi+2012, Biffi+2013,Biffi+2014,Biffi+2015,Cui+2016} and its contamination
   from AGN emission \citep{Biffi+2018b} by simulating the X-ray emission of the hot gas (thermal
   Bremsstrahlung with metal emission lines) and actively accreting SMBHs (intrinsically absorbed 
   PL) in the simulated clusters.
   Additionally, the general approach of the code allows for the treatment of different potential 
   X-ray sources within the simulation. The corresponding emission model, however,
   must be constrained from source properties within the simulation and included in \textsc{unit 1}.
   
   \subsection{Hot gas emission model}
   The synthetic X-ray emission for each hot-phase gas element in the simulations is modeled based on
   their intrinsic thermal and chemical properties (density,
   temperature, metallicity). A single-temperature thermal emission model with
   heavy element emission lines {based on the Astrophysical Plasma Emission Code (APEC)} \citep{APEC} is then assumed for every gas element making use
   of the implementation in \xspec{} \citep{XSPEC}.
   For further details on the implementation, we refer the reader to the description provided in
   \cite{Biffi+2012,Biffi+2013}.
   
   \subsection{AGN emission model} \label{ssec:AGN_model}
   Similarly, the synthetic X-ray emission for accreting black holes in the simulations is modeled by
   determining their accretion rate with respect to their Eddington accretion rate.
   In order to represent the effect of dense
   torus absorption, the X-ray spectrum is modeled using an intrinsically absorbed PL
   with column densities that are stochastically selected from the distribution given in \citet{Buchner+2014}.
   For further details on the implementation, we refer the reader to the description provided in
   \cite{Biffi+2018b}. The code has since been updated to model the AGN intrinsic absorption
   using the \texttt{TBABS} photo-ionization cross sections \citep{TBABS} instead of the 
   \texttt{WABS} cross sections \citep{WABS}, both implemented in \xspec{}.
   
   \subsection{XRB emission model} \label{ssec:XRB_model}
   Following the modeling of AGN and hot gas emission in the \phox{} code, we similarly
   determined the X-ray emission coming from a stellar resolution element of the simulation.
   We made use of the SSP properties determined in Sect. \ref{sec:SSPs} to calculate an expected
   number of point sources $\tilde{N}$ per stellar element. We then sampled the respective XLF
   based on the following recipe.
   We selected the correct XLF from a stellar age criterion: if the age of the SSP is
   $\tau < 30$ Myr, we classified it as HMXB hosting. If $\tau > 1$ Gyr we classified the 
   SSP as LMXB hosting. Both choices are based on expectations from binary evolution simulations
   \citep[see, e.g., the review by][]{Lewin+2006} and are well motivated from observations. In the HMXB case
   \citet{Shty+2007} found peak formation efficiency in the Small Magellanic Cloud
   at $\approx 50$ Myr, \citet{Garofali+2018} found peak formation efficiency in M33 at
   $\approx 40$ Myr, and \citet{Antoniou+2019} find peak formation efficiency again in the Small Magellanic Cloud
   at $\approx 30-40$ Myr, all by correlating the specific SFH of observed stellar regions with HMXBs.
   Since these values are dependent on the stellar models assumed for the observed regions
   we settled for the value given by the lifetime function (see Sect. \ref{sec:SSPs}). In the LMXB case
   the 1 Gyr boundary yields a donor mass limit of $1.75\,\Msun$, which reflects the Roche-lobe overflow scenario of accretion onto a CO.\par From the XLF we constructed a pseudo cumulative density function (pCDF).
       This requires setting a luminosity interval $[L_{\mathrm{min}}, L_{\mathrm{max}}]$
       at which the pCDF $C(L)$ will be defined:
       \begin{equation}
           C(L) = 1 - \dfrac{N(>L)}{N(>L_{\mathrm{min}})}\, , 
       \end{equation}
       with $N(>\!L)$ being the integrated XLF from step 1, and $C(L) \in [0,1]$. Typically,
       $L_{\mathrm{max}}$ is chosen to be the cutoff luminosity of the respective XLF.
       It should be noted that $C(L)$ is sensitive to the choice of $L_{\mathrm{min}}$ due to the PL
       dependence of the XLF. We then calculated $\tilde{N}$ from Eqs. \eqref{eq:tNLMXB} and \eqref{eq:tNHMXB}
       and chose $\lfloor\tilde{N}\rfloor$ uniformly distributed random numbers $p_i \in [0,1]$.
       If a separate uniform random number $\tilde{p}$ satisfies
       $\tilde{p}<\tilde{N}-\lfloor\tilde{N}\rfloor$, we draw an additional random number $p_i$.
       For each $p_i$ we determined a corresponding luminosity $L_i$, at which $C(L_i)=p_i$.
       From this we can calculate the total luminosity of the SSP as
       \begin{equation} \label{eq:Lssp}
           L_{\mathrm{SSP}} = \sum_i L_i\, .
       \end{equation}
   This recipe yields a SSP with a single X-ray luminosity in the same energy range 
   the XLFs have been defined in. As such, it will be treated as a single
   XRB-like point source. Specifically, we described XRB emission spectra as redshift-dependent,
   absorbed PLs of the form 
   \begin{equation} \label{eq:xrb_spec}
       A(E) = w(E(1+z)) \times K\left[ E\left(1+z\right)\right]^{-\Gamma}\, ,
   \end{equation}where $K$ is the spectrum normalization at $1\,\mathrm{keV}$ and $\Gamma$ is the photon
   index as given in the XSPEC manual (\cite{XSPEC}). Absorption was modeled after the
   \cite{TBABS} photo-ionization cross sections, and the model includes gas, grain, and molecule components in the ISM, which is expressed in Eq. \eqref{eq:xrb_spec} with $w(E)$.
   The redshift dependence of the spectrum allows for consistent modeling in a cosmological
   context. The total observed luminosity of the SSP (Eq. \eqref{eq:Lssp}) in a certain
   energy band $E\in[E_1,E_2]$ is then related to Eq. \eqref{eq:xrb_spec} by
   \begin{equation} \label{eq:Fssp}
       L_{\mathrm{SSP}} = F\int^{E_2}_{E_1} A(E) EdE\, ,
   \end{equation}
   with $F = 4\pi \mathcal{D}_L(z)^2\times 1.602\cdot10^{-9} \mathrm{erg\,keV^{-1}}$
   being the rescaling factor between flux and luminosity and $\mathcal{D}_L(z)$ the luminosity
   distance inferred from the underlying cosmology. The energy range $[E_1,E_2]$ has to be
   chosen in correspondence to the range in which the XLFs were defined, which is commonly
   adopted as $E_1 = 0.5\,\mathrm{keV}$ and $E_2 = 8\,\mathrm{keV}$ in the observed frame. We are then able to
   constrain the spectrum normalization $K_{\mathrm{SSP}}$ using Eqs. \eqref{eq:Lssp} and
   \eqref{eq:Fssp}, obtaining   \begin{equation}
       K_{\mathrm{SSP}} = \dfrac{\sum_i L_i}{F\int_{E_1}^{E_2}w(E)[E(1+z)]^{-\Gamma+1}dE}\, .
   \end{equation}
   
   A fully detailed description of XRB spectra is not possible in our modeling. For instance, the
   cosmological context of
   our approach makes it impossible to capture XRB variability on small timescales, so   changes in spectral hardness have to be accounted for statistically.
   Therefore, we opt for an average X-ray emission spectrum for the two XRB types,
   in line with typical values found in the literature.
   For our analysis, we chose the PL slope $\Gamma_{\mathrm{LMXB}} = 1.7$
   in the LMXB case and $\Gamma_{\mathrm{HMXB}} = 2$ in the HMXB case. Both slopes
   are motivated from observational data where the $\Gamma_{\mathrm{HMXB}}$ follows 
   assumptions of \citet{Mineo+2012a} and $\Gamma_{\mathrm{LMXB}}$ follows the median slope
   of all high-confidence XRBs in \citet{Lehmer+2019}. For the absorption component we
   assume the median column density $N^{\mathrm{xrb}}_H = 2\cdot10^{21}\,\mathrm{cm}^{-3}$ from the same sample of XRBs in
   \cite{Lehmer+2019}. This choice of parameters results in spectra resembling 
   typical XRBs in their low/hard-state \citep[see][]{Remillard+2006,Done+2007,Sazonov+Khabibullin2017c}.
   Additionally, it was found that $\Gamma_{\mathrm{LMXB}} \sim 1.4-1.8$ is sufficient to
   describe the hot atmosphere of EGs \citep{Boroson+2011,Wong+2014,Babyk+2018}.
   We do not include any other form of stellar X-ray sources such as active binaries or cataclysmic
   variables since their cumulative luminosity is at least an order of magnitude below the
   total XRB luminosity \citep[see, e.g.,][]{Boroson+2011,Babyk+2018}. Typically applied corrections for unresolved X-ray emission of cataclysmic
   variables and active binaries are $\sim 8\cdot 10^{27}\,\ulum\Msun^{-1}$ according to \citet{Babyk+2018}. For young stellar objects, \citet{Mineo+2012b} estimate a collective unresolved emission of $\sim 2\cdot 10^{38}\ulum$ per unit SFR.
   
   Having determined $K$, $\Gamma$ and $N^{\mathrm{xrb}}_H$ of a single SSP, we can compute the resulting
   XRB spectrum by adopting the \texttt{ztbabs} and \texttt{zpowerlw} model embedded in
   \texttt{XSPEC}, from which we calculated an ideal photon list associated with each SSP.
   
\section{Simulated data set}
\label{sec:data_set}
\begin{figure}
   \centering
   \includegraphics[width=\hsize]{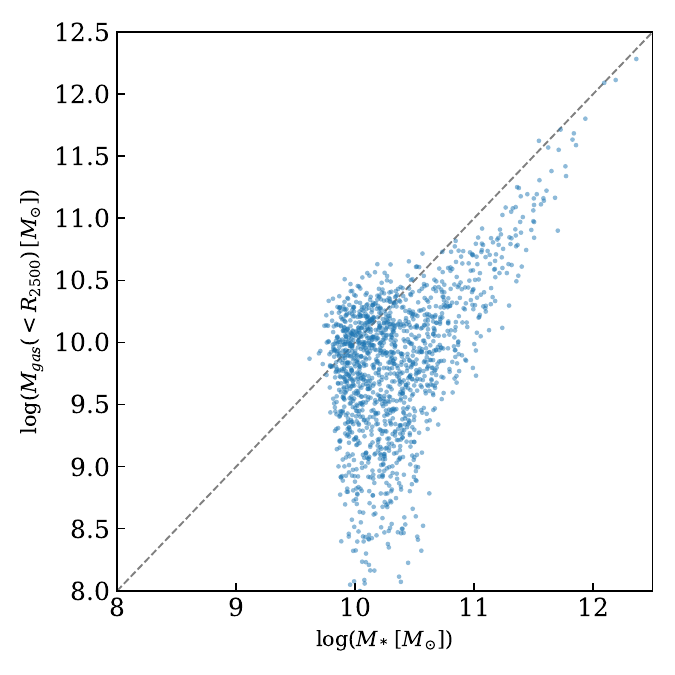}
      \caption{$M_{\mathrm{gas}}-M_*$ relation of our simulated galaxy sample. The dashed diagonal line indicates a one-to-one ratio.
              }
         \label{fig:mgas_mstr}
   \end{figure}

   For our analysis we extracted galaxy sized halos at redshift $z=0.07$ from the {Box4/uhr} cosmological
   volume of the {Magneticum Pathfinder simulation} set and artificially placed them at $z=0.01$.
   Assuming fiducial values for effective area $A_{\mathrm{eff}} = 1000\,\mathrm{cm}^2$ and exposure time
   $T_{\mathrm{exp}} = 10^5\,s$, we selected all halos generated by the \textsc{subfind} algorithm \citep{subfind2001,subfind2009}
   that have a total stellar mass $M_* > 10^{9.7}\,\Msun$ within a sphere of $R_{2500}$ around the halo center,
   which also includes a few massive group-like objects with stellar masses of $M_* \sim 10^{12}\,\Msun$.
   Halos with stellar masses below a threshold of $M_* < 10^{10}\,\Msun$ are resolution limited: the SPH 
   implementation fails to correctly reproduce the required star-formation for low-mass halos.
   We illustrate the $M_{\mathrm{gas}}-M_*$ relation in Fig. \ref{fig:mgas_mstr}. Halos that have high stellar masses
   and low gas masses were subjected to feedback processes and were depleted of their gas content, which is connected to resolved physical processes within the simulations.
   
   The selected set consists of $1480$ objects of which $335$ are considered actively star-forming,
   based on the limit $\mathrm{SFR} > 0.4\, \usfr$, with SFR derived from the total mass of
   stars born in the past 100 Myrs within the projected volume of each galaxy. This SFR estimator thus probes similar timescales of the SFH as typical tracers used in the literature such as in \citet{Mineo+2012a}. X-ray luminosities were calculated from the photon
   list obtained from a cylindrical volume around each galaxy with radius $R_{2500}$ and projected along the l.o.s. for a length of $2R_{2500}$. The l.o.s. direction coincides with the $z$-axis of the simulation box. We chose $R_{2500}$ as the smallest available scale invariant size from the halo catalog and to focus our analysis on the inner regions of selected galaxies in our sample. Reported comparison values probe projected radii on ISM scales consistently smaller than $R_{2500}$: \citet{Strickland+2004} and \citet{Mineo+2012a} within the $D_{25}$ isophotes, \citet{Bogdan+2013} within 0.05-0.15 $R_{200}$ (20-60 kpc), \citet{Lehmer+2016} within $\gtrsim 10$ kpc (see chapter on stacking procedure) and \citet{Lehmer+2022} within $\sim 50$ kpc.
    
\begin{figure}
   \centering
   \includegraphics[width=\hsize]{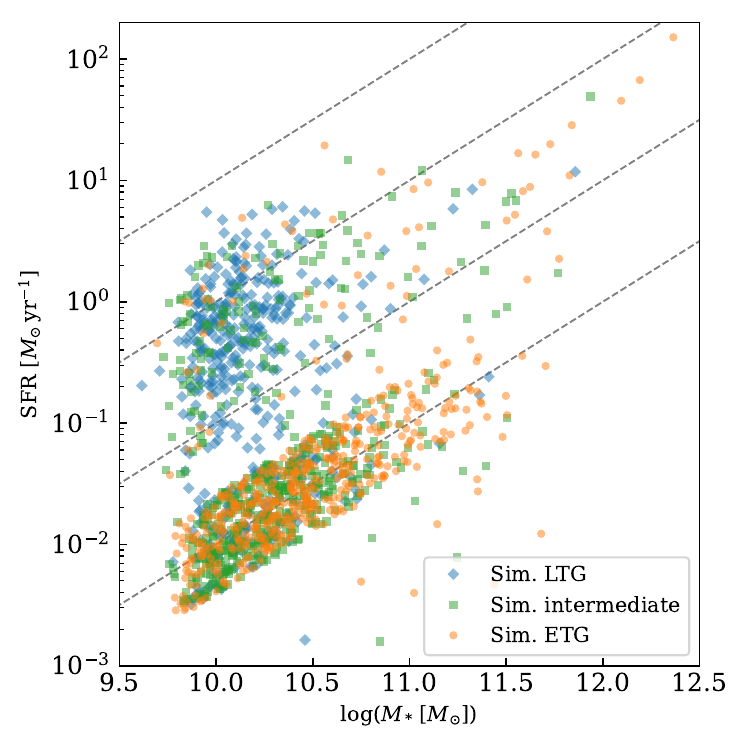}
      \caption{$\mathrm{SFR}-M_*$ relation of our sample. The dashed diagonal lines correspond to constant $\log \mathrm{sSFR}$ of -9 (top) to -12 (bottom) in increments of 1. Each galaxy was color-coded by its respective $b$ value \citep{Teklu+2015} with the classification scheme from \citet{Schulze+2020} (\textit{red}: elliptical, \textit{green}: intermediate, \textit{blue}: disk). Galaxies clustered at $\log \mathrm{sSFR} = -12$ were given a SFR estimate based on the \cite{Weinmann+2010} approach (see text).
              }
         \label{fig:sfr_mstr}
   \end{figure}

   Because \phox{} was originally conceived as a tool to study the X-ray emission of galaxy clusters, the
   calculated emission for the hot gas component in galaxies does not account for self-absorption from the ISM.
   We attempt to model self-absorption following observational derivations of intrinsic ISM emission by
   employing an additional \texttt{TBABS} model with $N^{gas}_H = 5\cdot 10^{21}\,\mathrm{cm}$ at the source redshift
   \citep[see, e.g.,][]{Mineo+2012b, Gilbertson+2022, Lehmer+2022}. In Fig. \ref{fig:sfr_mstr} we show our galactic sample on the
   SFR-$M_*$ plane color coded by the $b$ value, which is a measure of galaxy morphology from the intercept of
   $M_*-j_*$ relation  \citep{Romanowsky&Fall2012,Teklu+2015}. 
   We adopt the classification scheme by \citet{Schulze+2020} with $b>-4.35$ for disk galaxies (\textit{blue}),
   $-4.73<b<-4.35$ for intermediates (\textit{green}) and $b<-4.73$ for spheroids (\textit{red}).
   To increase the sample size of low-metallicity star-forming galaxies, we followed the approach
   taken by \cite{Weinmann+2010}, assigning a uniformly distributed $\log\,\mathrm{sSFR}$ in the range
   -12.4 to -11.6 for halos with $\mathrm{SFR}=0$. This can be seen in Fig. \ref{fig:sfr_mstr} as an apparent
   strip of mostly spheroidal galaxies located around the $\log\,\mathrm{sSFR} = -12$ line. We emphasize that galaxies distributed according to this criterion will not be used to derive hot gas properties in star-forming halos since the assigned SFR is not connected to gas properties within the simulations.
   
   Subsequent analysis of the generated ideal photons will concentrate on the unfolded photon lists without
   taking into account any instrumental response.

\section{Validation of the modeling}
\label{sec:Results}

   \subsection{Spatial distribution of XRB emission}
   
\begin{figure}
   \centering
   \includegraphics[width=\hsize]{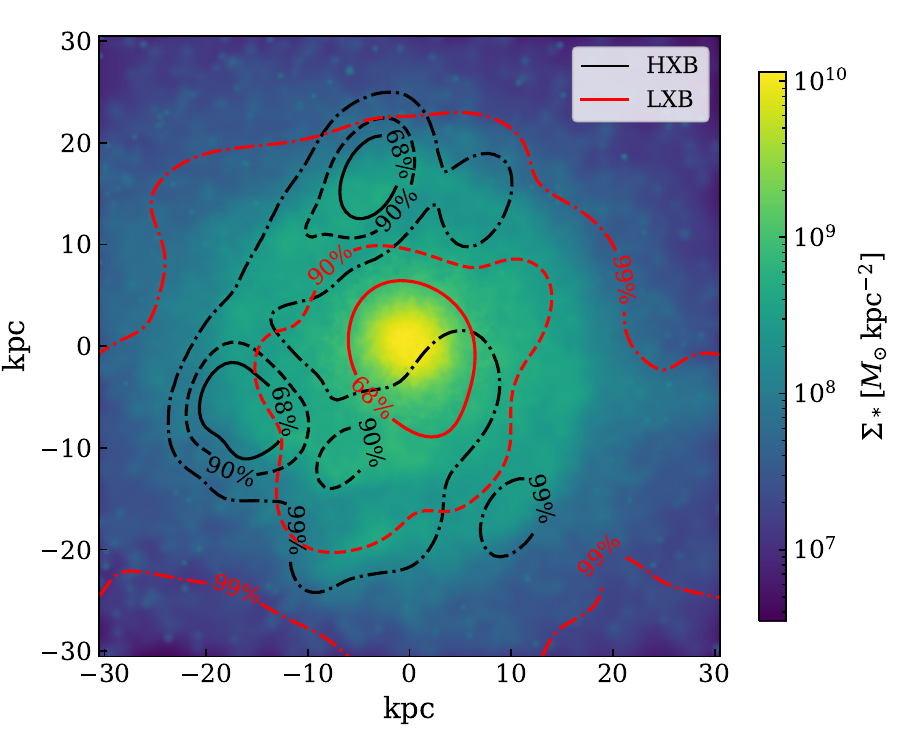}
      \caption{Projected stellar surface mass density ($\Sigma_*)$ of the poster-child disk
               galaxy in Box4 of Magneticum with halo ID {13633}. Contours indicate the 68th, 90th, and 99th percentiles of X-ray photon
               positions for HMXBs (\textit{black}) and LMXBs (\textit{red}).
               }
         \label{fig:xray_contour_mass}
   \end{figure}

\begin{figure}
   \centering
   \includegraphics[width=\hsize]{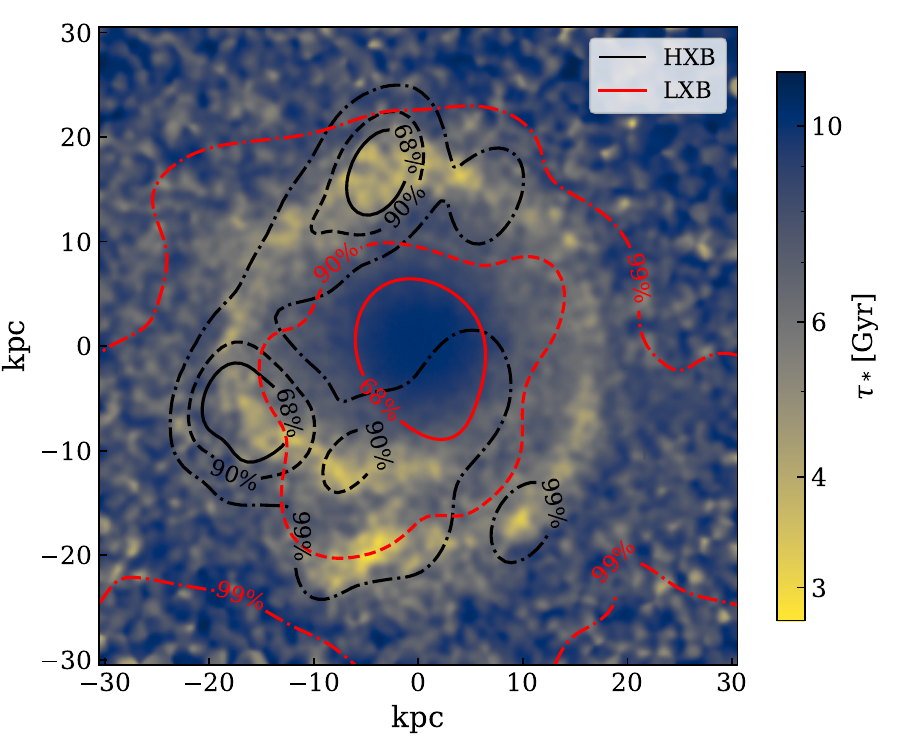}
      \caption{Mass-weighted stellar age ($\tau_*$) of the poster-child disk
               galaxy in Box4 of Magneticum with halo ID {13633}. Contours indicate the 68th, 90th, and 99th percentiles
               of X-ray photon positions for HMXBs (\textit{black}) and LMXBs (\textit{red}).
              }
         \label{fig:xray_contour_age}
   \end{figure}

   As a first step we show that the XRB seeding follows the correct spatial distribution of XRB emission. We selected a poster-child late-type galaxy from
   our sample with $M_* = 10^{11.3}\,\Msun$, $\mathrm{SFR}=8.4 \,\usfr$ and $R_{2500} = 91\,\mathrm{kpc}$ that appears face on to the line of sight. We focus our showcase to the inner 60 kpc in order to better capture
   the main stellar body of the galaxy.
   In Fig. \ref{fig:xray_contour_mass} we show the percentile contours of photons emitted by HMXBs (\textit{black}) and LMXBs (\textit{red})
   on top of the line-of-sight projected stellar mass map of the galaxy. As expected, the LMXB
   emission spans the whole field of the galaxy with its center over the galaxy bulge
   and decreasing with distance from the center reflecting the stellar surface density ($\Sigma_*$). 
   In contrast, HMXB emission is
   confined to the stellar mass overdensities surrounding the galaxy center.
   The reason for this can be seen in Fig. \ref{fig:xray_contour_age} where
   we show the same X-ray contours on top of the mass-weighted stellar age map of the galaxy.
   It shows that HMXB emission is bound to much younger stellar fields of the galaxy.
   In particular, the stellar age map shows the complex intersection of the two XRB components in the galactic field \citep[e.g.,][]{Lehmer+2017, Gilbertson+2022}.
   While this general behavior for the two XRB types is expected,
   the ability to clearly distinguish the origin of X-ray emission
   of a galaxy is a major advantage and enables predictions on theoretical aspects of X-ray spectra of
   unresolved galaxies in high redshift observations.
   We note that the spatial distribution of emitted photons was smeared out according to
   a Gaussian positioned at the projected particle location with standard deviation equal to the
   simulation intrinsic smoothing length of the SSP. Because of that, XRBs are not visible as point sources.
   
   \subsection{XLFs}
\begin{figure}
   \centering
   \includegraphics[width=\hsize]{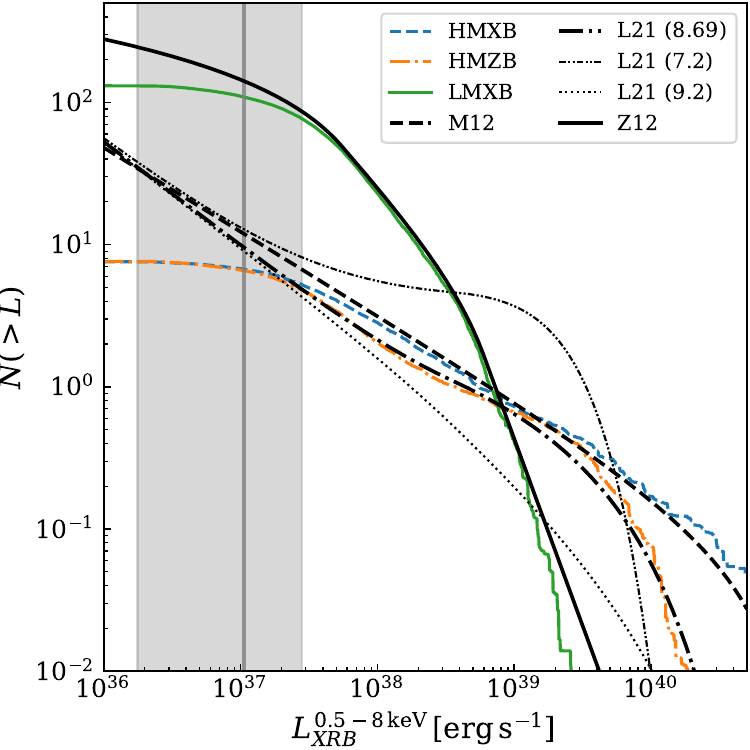}
      \caption{Integrated XLFs from Eqs. \eqref{eq:xlf_LMXB}, \eqref{eq:xlf_HMXB}, and \eqref{eq:xlf_hzb} normalized by their
      respective scaling (SFR and $M_*$). The \textit{dash-dot-dotted} line is the XLFs of the L21 model, for $\log[\mathrm{O/H}]+12 = 7.2$, and the \textit{dotted} line for
      $\log[\mathrm{O/H}]+12 = 9.2$. Colored lines are average XLFs obtained from the luminosity sampling in \phox{}: from \cite{Mineo+2012a} (M12) in \textit{dashed blue},
      \cite{Lehmer+2021} in \textit{dash-dotted orange} (L21), and \cite{Zhang+2012} in \textit{solid green} (Z12).
      Corresponding lines share the same line style. The \textit{gray} vertical line at $\log L \sim 37$ marks the maximum luminosity of a source
      emitting one single photon with an average energy of 0.5-8 keV. 
              }
         \label{fig:xlf}
   \end{figure}

   For the XLF reconstruction we chose a lower luminosity limit of $L_{\mathrm{min}} = 10^{35}\,\ulum$,
   since all XLFs are expected to have a turn-over at luminosities $\lesssim 10^{35}\,\ulum$. This is due to
   neutron star XRBs generally experiencing lower Eddington accretion rates than their black hole
   counterparts, as well as a strong magnetic field influencing the accretion flow
   \citep{Shty+2005a, Shty+2005b}. In addition, the integrated luminosity from the XLF will be dominated by the high luminosity end, so lower value choices of $L_{\mathrm{min}}$ will not change the integrated luminosity significantly. In Fig. \ref{fig:xlf}
   we compare the integrated XLFs (colored), reconstructed from the approach outlined in Sect. \ref{ssec:XRB_model}, with their
   original model (black). Corresponding XLFs are indicated with the same linestyle. 
   Luminosities of individual XRBs were calculated from the simulated photon lists associated with the SSP element hosting the XRB.
   In detail, each colored line is derived from the total photon box produced
   for each SSP in the simulation. The apparent turn-over of the reconstructed lines is
   unrelated to any physical effects associated with neutron star XRBs but can be explained by the
   following:
   (1.) low luminosity XRBs are unlikely to produce observable photons given the distance of the
   simulation box (low flux); and (2.) in cases where the XRB population size is greater than one per SSP,
   low luminosity XRBs are removed from the XLF through the summation process of
   Eq. \eqref{eq:Lssp}. The corresponding luminosity for a XRB emitting a single photon in our setup, assuming 
   an absorbed PL spectrum, is marked by the vertical gray band at $L_X \sim 10^{37}\,\ulum$, which
   roughly coincides with the turnover given the energy range of $0.5 - 8$ keV. The solid vertical gray line is the luminosity of a photon with mean energy $\sim 3\,\mathrm{keV}$ according to Eq. \eqref{eq:xrb_spec}.
   Thus, SSPs with $ L_{\mathrm{SSP}} < 10^{37.4}\,\ulum$
   will produce at most one single photon given the discrete photon sampling in \phox{}. 
   We note that our setup is not dependent on background flux calibration, since our goal in Unit 1
   of \phox{} is to create an ideal photon list.
   We find excellent agreement between the reconstructed XLFs (colored lines) and their original model (\textit{black}, same line style)
   above the single photon luminosity limit. The average reconstructed XLF for the L21 model (\textit{orange}) is
   consistent with solar values for $12+\log[\mathrm{O/H}]$, which is almost indistinguishable from the \cite{Mineo+2012a} XLF. Also shown is the L21 model at $12+\log[\mathrm{O/H}]=9.2$ and 7.2, which are not good descriptions of the average HZB XLF.
  
   \subsection{Scaling relations}
   Since our approach to assigning luminosities to SSPs is agnostic to its environment, that is, if the SSP is part of a star-forming halo
   within the simulation, we first verified the self-consistent emergence of well-known galactic X-ray scaling relations.
   
\begin{figure}
   \centering
   \includegraphics[width=\hsize]{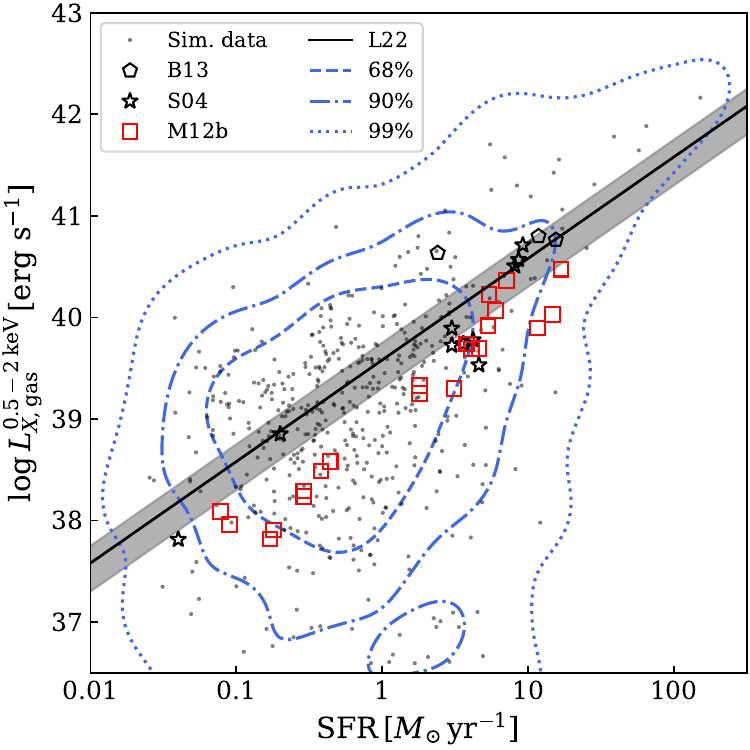}
      \caption{Relation between hot gas X-ray luminosity (0.5-2 keV) and the SFR for star-forming galaxies in the simulations.  We only show galaxies with $\log (\mathrm{sSFR}) > -11.6$, which
      removes all galaxies that fulfill the \citet{Weinmann+2010} criterion. This is done to reduce the influence of unresolved galaxies on the relation as well as remove weakly star-forming galaxies. 
      The \textit{solid black} line indicates the linear relation from \citet{Lehmer+2022}. The gray shaded area is the corresponding 1$\sigma$ standard deviation. Data points are bolometric luminosities of the diffuse emission taken from \cite{Strickland+2004} (SO4) in the 0.3-2 keV band, \cite{Mineo+2012b} (M12b) in the 0.5-2 keV band, and \cite{Bogdan+2013} (B13) in the 0.5-2 keV band.}
         \label{fig:lx_sfr_gas}
   \end{figure}

   Studies of the diffuse galactic X-ray emission revealed a linear relationship with SFR in the soft X-ray band, which is associated with ISM heating through SN remnants and stellar wind from heavy stars \citep{Ranalli+2003, Gilfanov+2004a, Mineo+2012b, Bogdan+2013}.
   However, it is observationally challenging to separate pure ISM emission from unresolved
   emission from point sources. In fact, \cite{Mineo+2012b} and \cite{Anderson+2013} show that
   two-thirds of the galactic extended emission in the soft band can be attributed to XRB emission,
   which they estimated from XRB scaling relations. Since \phox{} was designed to reproduce X-ray emission
   from galaxy clusters, we have to verify that galactic ISM properties are appropriately reproduced.
   In Fig. \ref{fig:lx_sfr_gas}
   we show the relationship between the hot gas X-ray luminosity and SFR in the soft X-ray band (0.5-2 keV).
   The blue contours enclose the 68th, 90th, and 99th percentiles of galaxies in our simulated sample also indicated by underlying green dots. We only show galaxies with $\log \mathrm{sSFR} > -11.6,$  which removes all galaxies falling in the \cite{Weinmann+2010} criterion. This is done to reduce the influence of unresolved galaxies on the relation as well as removing weakly star-forming galaxies.
   Also shown is best-fit linear relation from \citet{Lehmer+2022} with $\log (L_{X}^{0.5-2\,\mathrm{keV}}/\mathrm{SFR})=39.58^{+0.17}_{-0.28}$, using spectral energy distribution (SED) fitting to derive the SFH for each galaxy in their sample. 
   We include absorbed ISM luminosities from \cite{Mineo+2012b} (\textit{red} squares) with SFR derived from a combination of UV and IR tracers. Their reported values in the soft X-ray band are smaller compared to the L22 relation, which may be caused by differences in sample properties like metallicity. 
   They argue however, that their normalization might be higher by at least a factor of 2, due to uncertainties in observed column densities and depending on their fitting procedure.
   Additional bolometric gas luminosities are
   obtained from \citet{Bogdan+2013} (B13) (\textit{black pentagons}) in the 0.5-2 keV energy range and a radial range of
   0.05-0.15 $R_{vir}$ for late-type galaxies, with SFRs derived from the total IR luminosity. Although B13 aimed to study extended X-ray coronae
   around large spiral galaxies, the measured luminosities are consistent with a linear $L_X - \mathrm{SFR}$
   relation prompting a connection between SFR and galactic outflows.
   From \cite{Strickland+2004} (S04) we obtained data points for
   absorption-corrected total gas luminosities in late-type galaxies in the 0.3-2 keV energy range (\textit{black stars}), which also show a linear SFR relation. They derived SFRs from the far IR band luminosity.
   Assuming the same energy range of 0.5-2 keV and no absorption correction for S04 values, resulting luminosities would be slightly lower.
   Within the 90th percentile contours, our galaxy sample is consistent
   with observational data of the extended galactic emission. We notice, however, that while a linear trend exists in our data, the relation is less tight and contaminated by low-luminosity galaxies at $\mathrm{SFR \sim 1}$. Low-luminosity galaxies may have been the result of feedback processes removing large fractions of gas from the central halo, which we discuss in more detail in Sect.~\ref{sec:Discussion}.
   The intrinsic scatter in
   our data of $>1\,\mathrm{dex}$ is concerning, but might be a result of an uncurated sample.
   An important caveat is the limited resolution of the underlying simulation. Halos with stellar mass below the Milky Way stellar mass suffer greatly from the resolution restriction. In those cases, the total gas and stellar halo are composed of only a few hundred resolution elements, giving rise to a large intrinsic scatter due to the SPH implementation of the simulation as well as an eventual cutoff where a halo was not able to form stars (see Fig. \ref{fig:mgas_mstr}).
   
\begin{figure}
   \centering
   \includegraphics[width=\hsize]{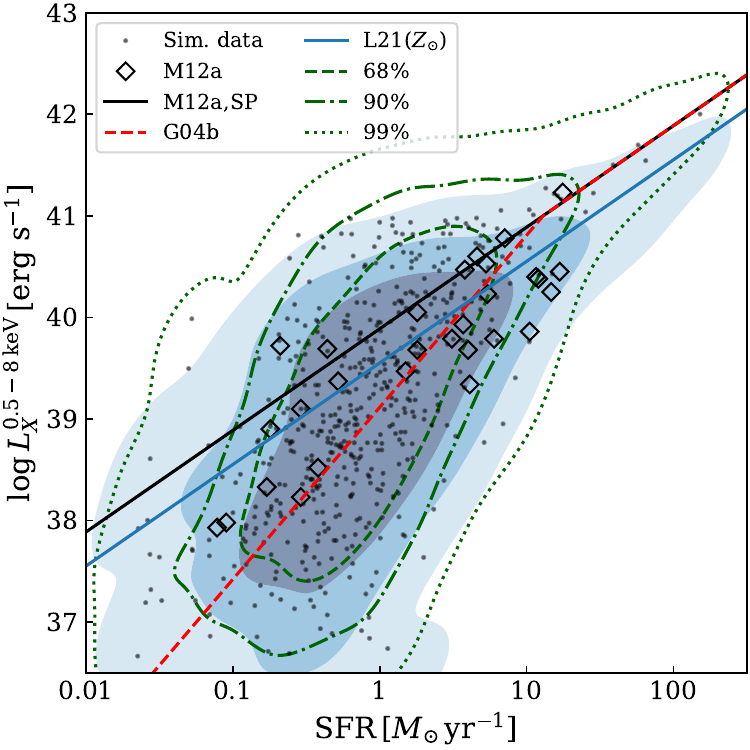}
      \caption{Same as Fig. \ref{fig:lx_sfr_gas}, but for HMXB luminosity in the 0.5-8 keV band. The filled \textit{blue} contours represent the same halos with HMXBs sampled from the L21 model (Eq. \eqref{eq:xlf_hzb}). The \textit{solid black} line is the expected relation from the employed HMXB model in Eq. \eqref{eq:xlf_HMXB}, and the \textit{dashed red} line is the predicted relation for discretely sampling data points from a PL distribution \citep{Gilfanov+2004b}. The \textit{solid} \textit{blue} line is the expected relation from the employed L21 model. Data points are taken from \cite{Mineo+2012a}.
              }
         \label{fig:lx_sfr_HMXB}
   \end{figure}

\begin{figure}
   \centering
   \includegraphics[width=\hsize]{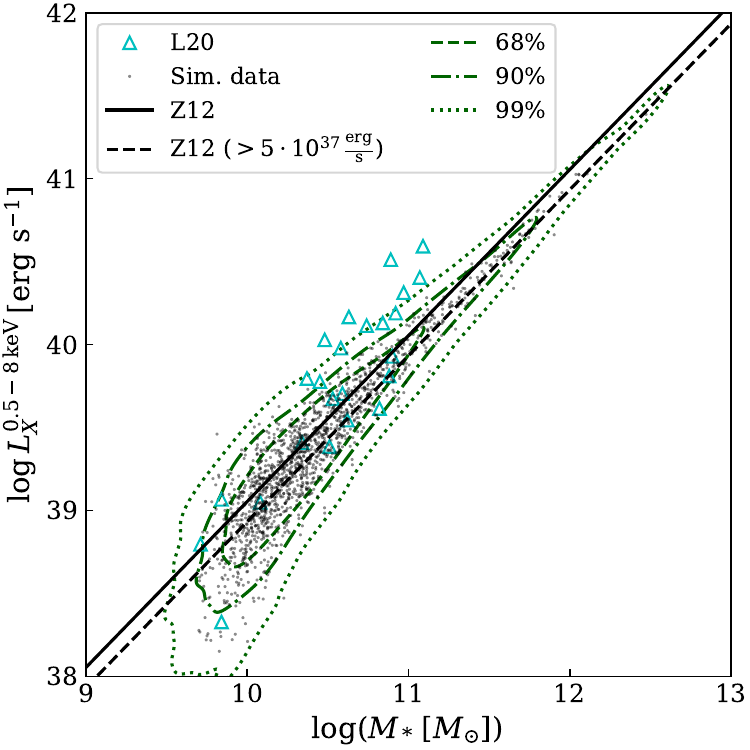}
      \caption{Same as Fig. \ref{fig:lx_sfr_HMXB}, but for LMXB emission in the 0.5-8 keV band. The \textit{solid black} line is the expected relation from the employed LMXB model in Eq. \eqref{eq:xlf_LMXB}. The \textit{dashed black} line is the same relation for a luminosity limit $L_{min}$ of $5\cdot 10^{37}\ulum$. Data points (\textit{cyan} triangles) are taken from \cite{Lehmer+2020}.}
         \label{fig:lx_mstar_LMXB}
   \end{figure}

   The primary goal of this paper is to model the stellar X-ray emission in the form of XRBs using data from cosmological simulations.
   In Figs. \ref{fig:lx_sfr_HMXB} and \ref{fig:lx_mstar_LMXB} we show the principal scaling relations for HMXBs and LMXBs, respectively.
   The contours have the same meaning as in Fig. \ref{fig:lx_sfr_gas}. The solid black line indicates the expected scaling relation
   from integrating the luminosity functions in Eqs. \eqref{eq:xlf_HMXB} and \eqref{eq:xlf_LMXB}. In the HMXB case in Fig.  \ref{fig:lx_sfr_HMXB}
   the reconstructed scaling relation is consistent with observations from \cite{Mineo+2012a} but shows significant deviation
   from the expected linear SFR dependence at low SFR values. This apparent deviation was noticed in observations of star-forming galaxies as well
   \citep{Grimm+2003, Gilfanov+2004a} and explained in \citet{Gilfanov+2004b} where the transition between linear and PL regime of the SFR relation is caused by low-number sampling of a single sloped XLF.
   The approximate relation calculated from
   \cite{Gilfanov+2004b} for the XLF in Eq. \eqref{eq:xlf_HMXB} is shown as the dashed black line and has a slope $\frac{1}{\gamma-1}$.
   For the LMXB case in Fig. \ref{fig:lx_mstar_LMXB}, the linear stellar mass dependence is well reproduced 
   but our data systematically underestimates the expected relation by a factor of $\approx 0.15$ dex. This is a manifestation of the one-photon limit associated with the XLF sampling. Also shown in Fig. \ref{fig:lx_mstar_LMXB} is the integrated luminosity function of Z12 (Eq. \eqref{eq:xlf_LMXB}) for $L_{min}=5\cdot 10^{37}\ulum$, which is approximately the upper bound of the one-photon luminosity limit. Accounting for the under-sampling of the LMXB XLF caused by the one-photon limit thus improves the relation. A similar effect is not seen in the HMXB since the under-sampled XLF only contributes less than 5\%.
   We do not observe a break in the LMXB-$M_*$ relation since our stellar mass limit ($M_* = 10^{9.7}\,\Msun$) is relatively high compared to the XLF normalization (Eq. \eqref{eq:xlf_LMXB}), which allows for sufficient sampling points. In the stellar mass regime of dwarf galaxies ($M_*\lesssim10^{9}$) we expect breaks in the $L_X-M_*$ relation as well \citep{Gilfanov+2004b}.
   
\begin{figure}
   \centering
   \includegraphics[width=\hsize]{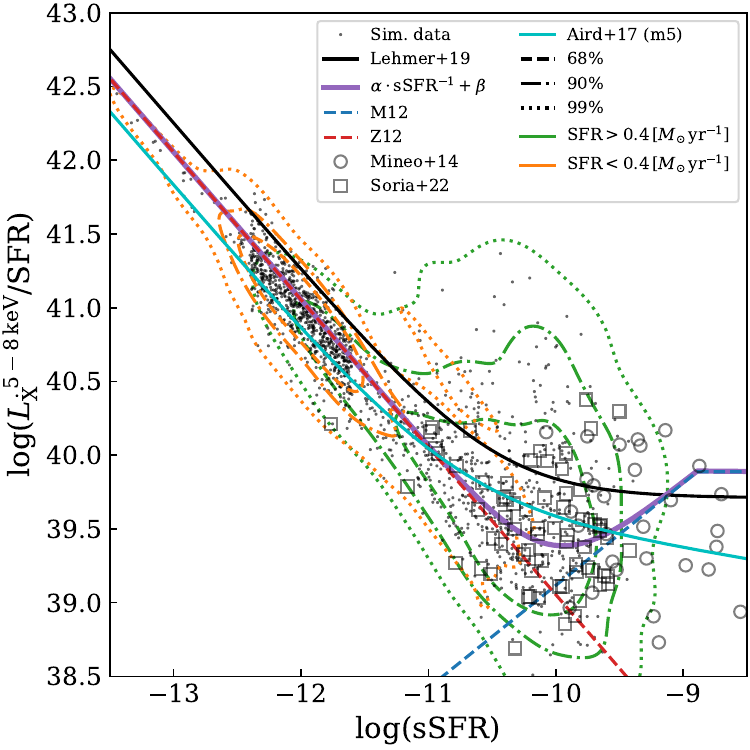}
      \caption{Relation between SFR-normalized total X-ray luminosity (both XRB types) and sSFR, for star-forming galaxies in Box4of Magneticum (contours: percentiles). The \textit{solid black} line is the relation of observed normal galaxies at $z\sim 0$ from \cite{Lehmer+2016}. The \textit{dash-dotted} line represents the relation for a set of local galaxies from \cite{Lehmer+2019}. The \textit{solid cyan} line is the relation obtained from model 5 (m5) in \citet{Aird+2017}, where we fixed $z=0.01$. The \textit{dashed red} line depicts the expected scaling relation for a pure LMXB contribution given our XLF choice. The \textit{dashed blue} line shows the expected broken SFR scaling relation of a pure HMXB contribution. Data points are taken from \cite{Soria+2022} (S22) and \cite{Mineo+2014} (M14).
              }
         \label{fig:lx_Lehm16}
   \end{figure}

   When combining the SFR-normalized emission from HMXBs and LMXBs, we expect a sSFR-dependent relation
   of the form 
   \begin{equation} \label{eq:Lehm16}
       L_X/\mathrm{SFR} = \alpha\,\mathrm{sSFR}^{-1}+\beta
   ,\end{equation}
   where $\alpha$ contains the linear mass
   dependence of the total LMXB luminosity and $\beta$ contains the linear SFR dependence of
   the total HMXB luminosity \citep{Lehmer+2010,Lehmer+2016,Lehmer+2019}.
   In Fig. \ref{fig:lx_Lehm16} we show the sSFR dependence of the SFR-normalized combined XRB luminosity of our sample with contours as in Fig. \ref{fig:lx_sfr_gas}.
   Additionally, we show data points from
   \cite{Mineo+2014} (M14) (SFRs as in M12a) from nearby galaxies and a study performed on star-forming galaxies within
   the Virgo cluster by \cite{Soria+2022} (S22) with SFR values derived from the $12\,\mu m$ WISE W3 measurement.
   We include an observationally derived relation (\textit{black, dash-dotted}) for local galaxies from \cite{Lehmer+2019} (L19) where SFRs were derived from a combination of far UV and 24 $\mathrm{\mu m}$ maps.
   The red dashed diagonal line represents the theoretical contribution from LMXBs given our model
   choice (Fig. \ref{fig:lx_mstar_LMXB}) and the blue dashed line corresponds to contributions from HMXBs
   (Fig. \ref{fig:lx_sfr_HMXB}). Because of the HMXB scaling relation being more similar to a BPL, the
   combined luminosity of LMXBs and HMXBs, normalized by SFR, takes a different shape (\textit{dashed purple})
   compared to the L19 relation: 
   it reaches a minimum value at $\mathrm{sSFR}\sim -10$ after which it begins to approach the linear HMXB regime.
   Our sample is in agreement with the purple relation and is broadly consistent
   with the relations from L19 for $\log \mathrm{sSFR} \lesssim -10.5$.
   However, due to the lack of a high sSFR halos in the simulated sample, we are not able to reliably demonstrate
   the expected turnover predicted by our model in purple.
   Interestingly, we find remarkable agreement between the simulated data and S22 data obtained from star-forming galaxies in the Virgo cluster. Furthermore, data from M14 connects seamlessly to both the S22 sample
   and our sample. We also compare model 5 of \citet{Aird+2017} (\textit{solid cyan}) in which they quantify the total $L_X$ of a star-forming galaxy as a function of $M_*$, SFR and redshift. They take into account their whole $0.1 < z < 2$ sample of galaxies and convert the total count rate of each galaxy from the 0.5-2 keV band to the 2-10 keV band using a constant conversion factor based on an absorbed PL with $\Gamma = 1.9$. It takes the form
   \begin{equation}
       L_X = \alpha(1+z)^{\gamma} M_* + \beta(1+z)^{\delta}\mathrm{SFR}^{\theta}\, ,
   \end{equation}
   with $\log\alpha[\ulum] = 28.81$, $\log\beta[\ulum] = 39.5$, $\gamma = 3.9$, $\delta = 0.67$, $\theta = 0.86$, $M_*$ in $\Msun$ and SFR in $\ulum$. We fixed the redshift to $z=0.01$ in line with our sample.
   Due to the nonlinear SFR dependence, model 5 falls off for higher SFR values. From a visual inspection alone, both our broken SFR scaling relation and model 5 appear to adequately describe the SFR-normalized luminosities from S22 at high sSFR.
   However, model 5 underpredicts luminosities in the low sSFR regime compared to our simulated data.
   In their study S22 show that most of their data points with $\log\mathrm{sSFR}$ lie below the median tracks of Monte-Carlo simulations of L19 for various populations of galaxies. They argue that inclination effects and differences in the used galaxy population are causing this discrepancy.

   \subsection{Metallicity relation of HMXB}
\begin{figure}
   \centering
   \includegraphics[width=\hsize]{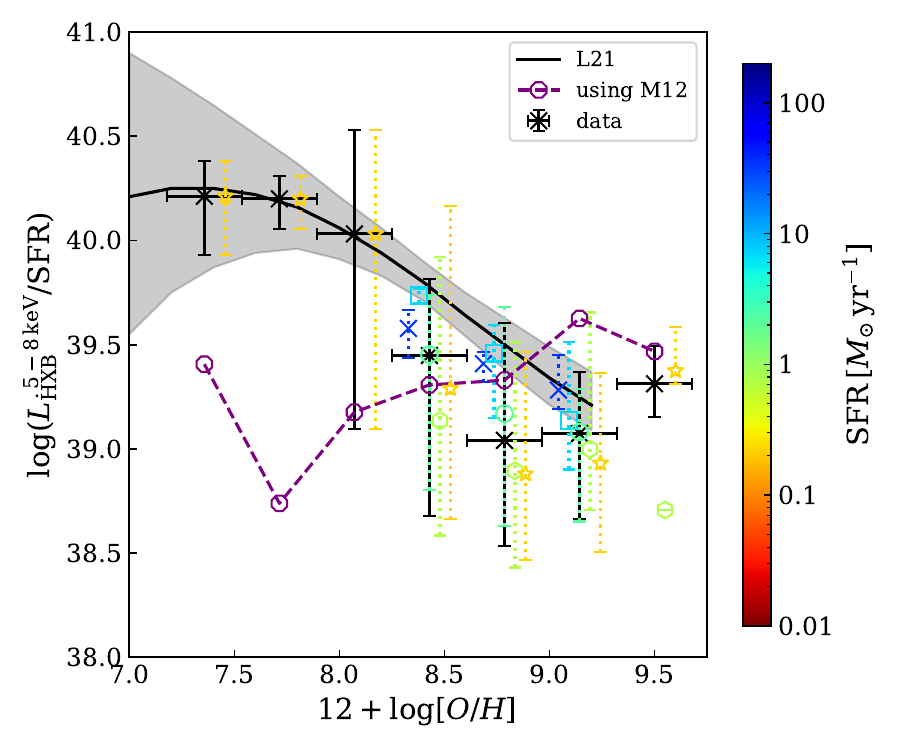}
      \caption{Metallicity dependence on the linear SFR scaling relation of HMXB luminosity. Oxygen fractions are derived from mass-weighted stellar metallicities of SSPs younger than 30 Myr. The \textit{black solid} line is the mean global model calculated by L21 in the limit of perfect luminosity function sampling. The shaded \textit{gray} area is its 16th-84th percentile margin. Our data are binned by oxygen fraction and SFR; SFR bins are indicated by color. Vertical error bars correspond to the 25th-75th percentile margin of the data within each metallicity bin, and horizontal \textit{black} bars indicate the width of the metallicity bin. Colored symbols corresponding to the SFR bins are displaced horizontally with respect to the metallicity bin center for illustrative purposes. The dashed \textit{purple}  line is the corresponding M12a model where no explicit metallicity dependence is assumed.
              }
         \label{fig:lx_Z_hzb}
   \end{figure}

   Numerous studies on the total HMXB emission in individual galaxies associate an increase in HMXB emissivity per unit SFR with a lower-metallicity
   environment of the HMXB. (\cite{Vulic+2021, Saxena+2021, Lehmer+2021, Garofali+2020, Fragos+2013b}). In this section we try to verify if our
   modeling allows us to capture the metallicity dependence of the HMXB luminosity function proposed by \cite{Lehmer+2021} (L21) (see Eq. \eqref{eq:xlf_hzb}). They obtained SFRs similar to L19 using far-UV and 24 $\mathrm{\mu m}$ maps and gas-phase metallicities from emission-weighted oxygen line ratios.
   In Fig. \ref{fig:lx_Z_hzb} we show the SFR-normalized X-ray luminosity of our star-forming galaxy sample against the metallicity derived from
   the mass-weighted oxygen fraction of the young stellar population ($< 30$ Myr). We binned our sample according to metallicity and SFR with horizontal bars representing the
   bin width in metallicity and vertical bars representing the 25th-75th percentile in luminosity.
   Black crosses are the median values for the whole star-forming sample while colored symbols correspond to
   different SFR bins. The SFR bins are displaced with respect to the metallicity bin center for illustrative purposes.
   Our full sample (\textit{black crosses with error bars}) is consistent with the theoretical model and shows the expected increase in luminosity for lower
   metallicities. Similarly, the high SFR sample ($\gtrsim 5\,\usfr$, blueish colors) is also consistent with
   the global model up to a oxygen fraction of 8-8.5. Our sample did not include highly star-forming galaxies
   with lower metallicities. For the sample at $\mathrm{SFR}\sim 1\,\usfr$ (green) we find lower median luminosities
   compared to the global model but are consistent within the 25th percentile margin. From the Markov chain Monte Carlo simulation performed in L21, it is however expected that the median luminosity
   decreases with lower SFR values due to worse HMXB count statistics. Despite this, the median luminosity for the
   low SFR sample (\textit{orange}) lies more than an order of magnitude above the relation predicted in L21
   for $\mathrm{SFR}\lesssim0.1$.
   Given our approach to retrieve HMXB luminosities for SSPs in the simulation, we find that, with the given mass
   resolution of the simulation, a single eligible SSP in a low SFR galaxy might already overcount the expected number of HMXBs. Additionally, only luminous SSPs will be sampled by \phox{} (see Fig. \ref{fig:xlf}).
   Since we did not
   take into account galaxies that do not have HMXB emission despite being star-forming, we are biased toward
   a low SFR sample with comparatively large luminosities and too large HMXB population.

\section{Relative emission of hot gas and XRBs} \label{sec:Counts}

   In this section we used the ideal photon list generated with \phox{} to obtain relative count ratios
   between the gas and XRB component in our galaxy sample. Additionally, we built average
   spectra of normal galaxies by first looking at the whole sample and subsequently dividing the
   sample according to different galactic properties. Average spectra are obtained by first
   normalizing each galaxy spectrum with its total emission in the 0.5-10 keV band and afterward
   taking the average in each energy bin of the stacked spectra.
   
   \subsection{Relative count ratio}
\begin{figure}
   \centering
   \includegraphics[width=\hsize]{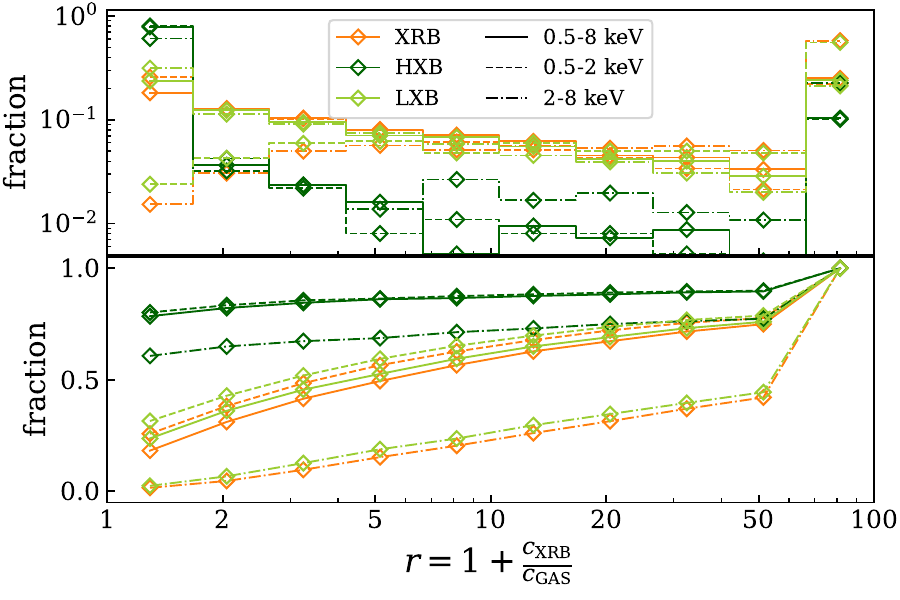}
      \caption{X-ray photon count ratio between the XRB and the hot gas component including all galaxies in our sample. Colors represent the component and line styles the energy range in the observed frame. The label ``XRB'' shows the sum of HMXB and LMXB emission. \textit{Top:} Differential fraction of galaxies with a given count ratio, $r$. \textit{Bottom:} Cumulative fraction of galaxies. Here, a ratio of $r\geq2$ means that the XRB component is dominant. Ratios exceeding the given range are collected in the last bin.
              }
         \label{fig:cnt_ratio}
   \end{figure}

   In order to investigate the relative contribution of the hot gas and XRB component
   in our galaxy sample, we compared the respective photon counts coming from each component.
   We split the contribution into different energy bands (0.5-2 keV, 2-8 keV, and 0.5-8 keV)
   and binned our galaxies according to the count ratio, $r$, which is defined as
   \begin{equation}
       r = 1+\dfrac{c_{XRB}}{c_{GAS}}\, ,
   \end{equation}
   where $c_{i}$ stands for the number of photon counts from the respective component.
   A value of $r>2$ indicates a dominant XRB component in our setup. In Fig. \ref{fig:cnt_ratio}
   we show the differential and cumulative fraction of galaxies in our sample with respect to the ratio $r$.
   We show the LMXB contribution, the HMXB contribution,
   and the summed contribution of LMXBs and HMXBs. Different line styles correspond
   to the different energy bands. Regarding the combined
   contribution of both XRB types we notice that $\sim90\%$ of our galaxies are dominated by XRB
   emission in the hard X-ray band (2-8 keV). Interestingly, this high fraction is almost
   exclusively caused by the LMXB emission. One has to keep in mind, however, that HMXBs only
   occur in galaxies with sufficient star-formation. Thus, the low HMXB contribution is rather an
   expression of the rareness of star-forming galaxies compared to non-star-forming ones in our sample.
   If we include the soft band (0.5-2 keV), where the hot gas emission is stronger, only
   $\lesssim40\%$ of our galaxies are still dominated by XRB emission. 
   
   \subsection{Average galaxy spectra}
\begin{figure}
   \centering
   \includegraphics[width=\hsize]{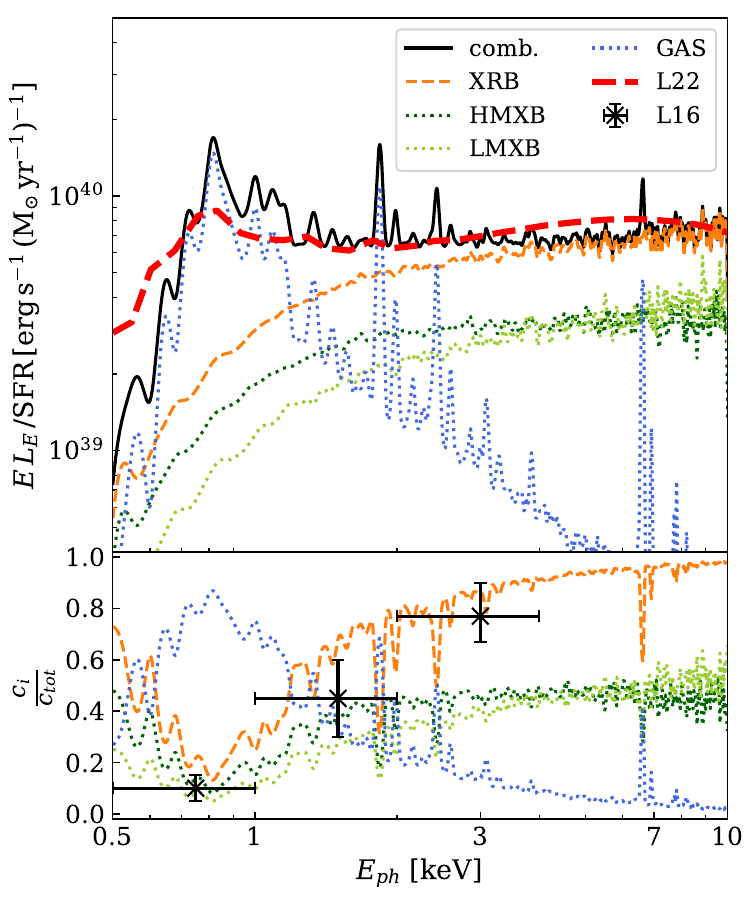}
      \caption{SFR-normalized average X-ray spectrum of our star-forming sample ($\mathrm{SFR} > 0.4\,\usfr$). The combined spectrum (\textit{black}) is made up of the HMXB (\textit{dark green}), the LMXB (\textit{lime green}), and the hot gas (\textit{blue}) component. The combined XRB emission is given in \textit{orange}. The \textit{red} line is the L22 average spectrum derived from observations of dwarf galaxies.
              }
         \label{fig:avg_spec}
   \end{figure}

   In addition to comparing the total counts received from each component, we construct an average
   spectrum from star-forming galaxy sample and quantify the contribution of each component in each energy
   bin. To do this we first normalize each galaxy spectrum to its 0.5-10 keV
   emission and taking the average in every energy bin. The results can be seen in Fig.
   \ref{fig:avg_spec}, which shows the total average spectrum, the average combined spectrum of both XRB types, and the average spectrum of hot gas. In the bottom panel of Fig. \ref{fig:avg_spec} we show
   the relative contribution of each component toward the total spectrum as a function of energy. Additionally,
   we added the average spectrum from \citet{Lehmer+2022}, which was derived for low-metallicity dwarf galaxies (\textit{dashed red}) and values for the relative XRB contribution in different energy bands at redshift $z=0$ from \citet{Lehmer+2016} (L16) in the lower panel of Fig. \ref{fig:avg_spec}.
   From a visual inspection, the combined average spectrum is similar to the L22 spectrum in the energy range 0.8-10 keV despite differences in mass and metallicity regimes in their sample. Below 0.8 keV our average
   spectrum falls off more steeply than the L22 spectrum. This is connected to differences in assumptions made for
   the different component spectra in L22. They use a two temperature APEC model with intrinsic absorption column
   densities similar to our $N^{gas}_H = 5\cdot10^{22}\,\mathrm{cm^{-3}}$ but use a weaker absorption coefficient
   $N^{\mathrm{xrb}}_H = 6\cdot10^{20}\,\mathrm{cm^{-3}}$ compared to our $2\cdot10^{21}\,\mathrm{cm^{-3}}$.
   Because of this, the XRB contribution is much higher for low energies. The sensitive behavior of the spectrum at low energies can also be seen when comparing the relative contributions of XRB with values from L16 in the
   lower panel. In L16 they build an average spectrum for massive star-forming galaxies from observations by \citet{Wik+2014}, \citet{Lehmer+2015}, and \cite{Yukita+2016}. While we predict strong contributions of XRBs in the 0.5-1 keV energy range with >30\%, L16 find much lower values of $\sim 10\%$. Our predictions agree well with L16 for harder energy bands (>1.5 keV). Interestingly, \citet{Wik+2014} also find high XRB contributions in the 0.5-1 keV band that are similar to our values, again highlighting the strong model dependence of the XRB contribution in the soft X-ray.
   Given our simple choice for the XRB emission model in the form of a single absorbed
   PL, we fail to capture the expected steepening of the XRB PL for $E>6\,\mathrm{keV}$
   that is predicted by \cite{Lehmer+2016} for normal galaxies, by population
   synthesis performed in \cite{Fragos+2013b}, and by, for example, \cite{Persic+2002} for star-burst galaxies.
   A more sophisticated model of the XRB spectrum would be needed to fully reproduce
   observed spectral shapes. 
   
\begin{figure*}
   \resizebox{\hsize}{!}
            {\includegraphics{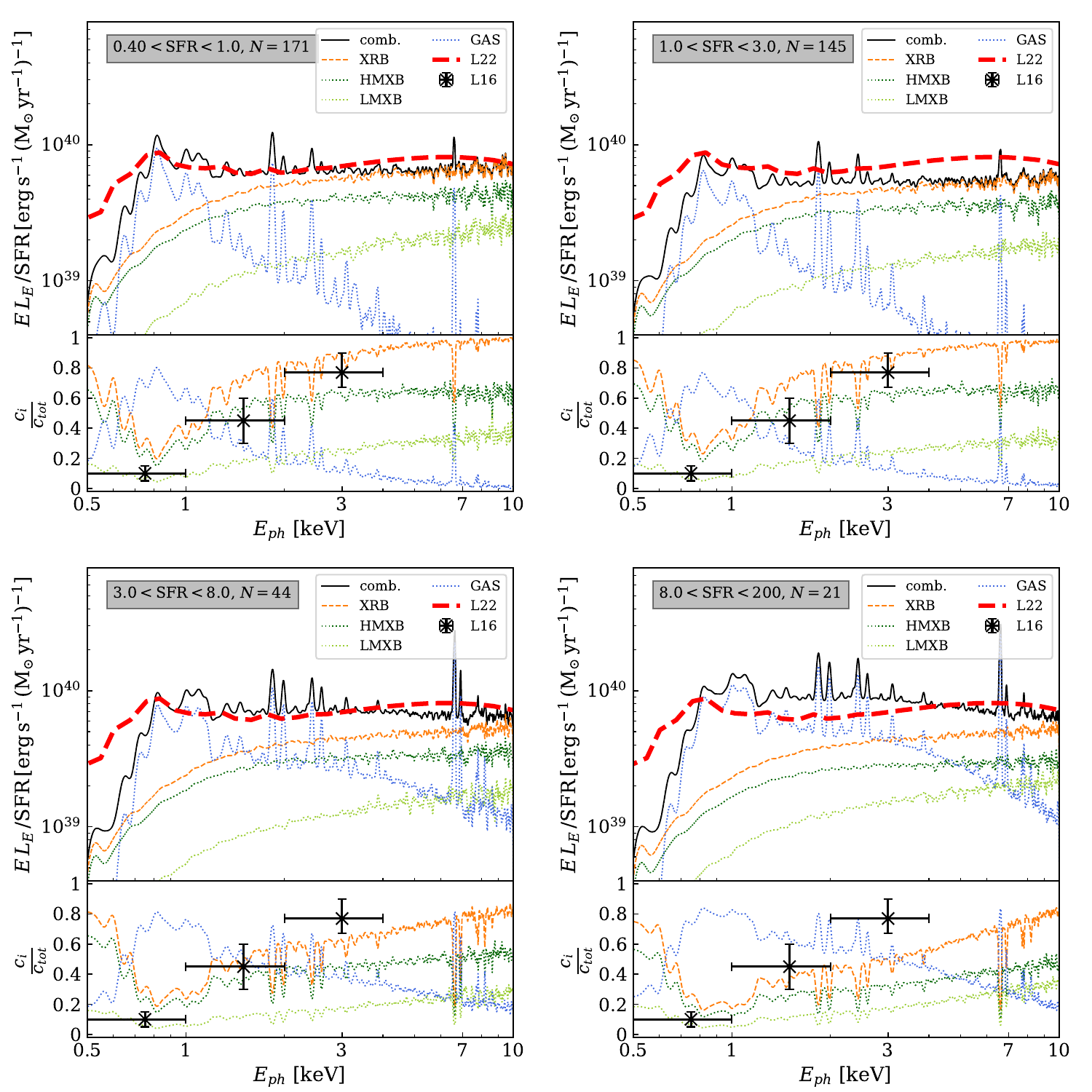}}
      \caption{Mean X-ray spectrum (\textit{solid black}) of our star-forming galaxy sample binned by SFR from lowest (\textit{upper left} panel) to highest (\textit{bottom right}) in the energy range 0.5-10 keV. The mean spectrum was calculated by normalizing each galaxy spectrum to its total 0.5-10 keV emission and taking the average in each energy bin. \textit{Dotted} lines represent the different components that contribute to the total spectrum: hot gas emission (\textit{blue}), LMXB (\textit{lime green}), and HMXB (\textit{dark green}). ``XRB'' (\textit{orange}) stands for LMXB+HMXB emission. The \textit{red} line is the L22 average spectrum derived from observations of dwarf galaxies. The lower part of each panel shows the fractional contribution of each component to the mean spectrum together with the XRB count ratios from L16. The gray box highlights the bin size and the number of galaxies, $N$, within each bin.
              }
         \label{fig:spec_contr_sfr}
   \end{figure*}

   Further disentanglement of the sample into increasing SFR bins yields a more refined view on the
   underlying galactic properties influencing the spectral composition. In Fig. \ref{fig:spec_contr_sfr}
   we show the same decomposition of galactic spectra as in Fig. \ref{fig:avg_spec} for four SFR bins.
   While the average XRB contribution stays consistent with previously discussed values, we see slight variations between each SFR bin. Specifically, in the lowest SFR bin (\textit{top left}) we notice a very strong XRB component that generally exceeds estimates from L16. By comparing to Fig. \ref{fig:lx_sfr_gas}, where we found galaxies with $\mathrm{SFR}\lesssim1\,\usfr$ having low $L_X$ within the 99th percentile contour, we attribute the strong XRB component to under-luminous gas-poor galaxies in the probed SFR regime and normalization issues when determining $\tilde{N}_{\mathrm{HMXB}}$ from Eq. \eqref{eq:tNHMXB}, which leads to more HMXBs per unit SFR. In the next higher SFR bin (\textit{top right}) the normalization of the spectrum drops slightly, mainly in the HMXB component because the normalization issue no longer applies here. Simultaneously, the hot gas component increases normalization in the hard band due to more massive galaxies contributing to the bin with hotter gaseous atmospheres. The relative contribution of the XRB component is consistent with L16 values in this case in the hard band. Further increasing SFR in each following bin yields even more prominent hot gas contributions in the hard band (see \textit{bottom left} and \textit{bottom right}). While the SFR-normalized total XRB luminosity stays almost constant in the two last bins, the XRB contribution is diminished by strong contributions from the hot gas atmospheres around massive, BCG-like galaxies with their deep gravitational potential. In these cases, the L16 values become less suitable to compare against since the considered samples have vastly different properties. In the highest SFR bin (\textit{bottom right}) the inclusion of group-like mass regimes also disfavors the comparison against the L22 spectrum that was derived using high redshift analogs for normal galaxies. 

   Since the two lowest SFR bins present the largest contribution in terms of sample number,
   the total average spectrum from Fig. \ref{fig:avg_spec} will mostly resemble to the upper panels of Fig. \ref{fig:spec_contr_sfr}.
   An important question associated with galactic X-ray spectra is their
   influence on the cosmic X-ray background. Upcoming planned X-ray missions such as Athena or the Light Element Mapper are sensitive enough
   to receive significant contributions from galactic emission to their background calibration.
   We plan to investigate this issue in an upcoming study.
   
\section{Discussion} \label{sec:Discussion}
\subsection{Gas luminosity scaling relation}
Compared to observational data we found that the $L_X-\mathrm{SFR}$ relation in the 0.5-2 keV band for the hot gas
component has a large intrinsic scatter of $\sim 1 \,\mathrm{dex}$ at $\mathrm{SFR} \lesssim 1$ with
observers typically reporting lower luminosities of $L_X \sim 10^{39}\times\mathrm{SFR}$ with
significantly smaller scatter \citep{Strickland+2004,Mineo+2012b,Bogdan+2013}. We propose six explanations
as to why we expected the scatter to be larger for the hot gas component.

First, in its core, the \phox{} code was perceived as a tool to study X-ray emission from galaxy clusters with relatively hot atmospheres of $T > 10^7\,\mathrm{K}$ \citep{Biffi+2012}. Thus, an intrinsic dusty absorption component was not required to realistically model cluster emission. In order to emulate a dusty absorber intrinsic to each selected galaxy of our data set, we applied a constant absorption model at the redshift of the selected halo consistent with ISM absorption column densities. While it improved the global relation compared to when no absorption is applied, it is not clear whether a strong absorption component is suited for halo outskirts beyond $0.5\,R_{2500}$. Furthermore, numerous studies explicitly mention that intrinsic absorption is highly dependent on the observed galaxy and may vary significantly \citep[e.g.,][]{Mineo+2012b, Lehmer+2016, Lehmer+2019, Lehmer+2022}.

Second, in several galaxies with elevated $L_X$ we found isolated pockets of dense hot gas clumps, despite specifically filtering out overheated star-forming gas elements from the simulation. This issue has to be investigated further but may be connected to the chosen projected volumes, which we explain in one of the following points.

Third, recent studies have shown that the extended hot gas atmosphere can significantly impact the total observed luminosity depending on the chosen projection radius \citep{Nica+2021,Comparat+2022}. In \citet{Comparat+2022} they studied the X-ray emission of the circumgalactic medium for halos of stellar mass $2\cdot10^{10}-10^{12}\,\Msun$ for different projection radii (80 kpc, 300 kpc). They further differentiated between star-forming and quiescent galaxies. They found that the enclosed luminosity in the 0.5-2 keV band increases by a factor of at least 2 for star-forming galaxies and $>10$ for quiescent galaxies if the enclosing projected radius is increased from 80 kpc to 300 kpc. In addition, they find that while quiescent galaxies clearly have an extended hot halo, star-forming galaxies share radial profiles with an expected diffuse unresolved component and weak AGN. This may indicate that, in order to more properly study gas scaling relations for the ISM, the projected radius has to be chosen more carefully.

Fourth, regarding the influence of AGN toward ISM properties, the Magneticum simulations include a thermal feedback model for SMBH described in \cite{Hirschmann+2014}. While SMBH properties are tracked by the simulations we did not specifically check for AGN activity in each of our selected halos. It is however known, that AGN experience different behavior compared to normal galaxies with respect to their X-ray radial profiles and scaling relations (\cite{Comparat+2019,Vulic+2021}). A proper disentanglement of AGN contribution to these relations goes beyond the scope of this paper. We prefer to more thoroughly analyze their effects in a future study.

Fifth, another influencing factor for gas properties in the ISM are ongoing mergers between galaxies. In some instances we found that our selected halos had equally massive companions within $R_{2500}$, possibly introducing an additional heating source through merger triggered interaction. We did not exclude merging halos from our sample since we selected for a good representation of different classes of objects for our XRB model. 

Finally, the most driving factor for uncertainty, especially toward low $L_X$, is probably the resolution limit of the total gas mass in the cosmological simulation. SPH simulations usually derive hydrodynamical quantities by interpolating between neighboring SPH elements with a set number of neighbors. As soon as a halo gas mass reaches below a lower mass limit, hydrodynamical effects like star-formation can no-longer be consistently resolved since the number of SPH elements in the halo becomes smaller than the needed neighbor number. The scatter is then dominated by the stochastic process with which sub-grid star-formation is handled in the implementation \citep[see][]{SH2003}.

We want to emphasize again that the hot gas component was not a major focus of this paper. Nonetheless, our approach shows that the simulation can recover a linear trend between gas X-ray luminosity and SFR if we select sufficiently resolved galaxies. However, the large scatter of over $\sim 1\,\mathrm{dex}$ shows that a simple approach to model the ISM is not optimal and needs more careful investigation.

\subsection{XRB scaling relations}
The main purpose of this paper was to introduce the viability of a semi-analytic approach to model X-ray emission from XRBs using hydrodynamical cosmological simulations. We connected locally derived luminosity functions of XRBs and their intrinsic behavior to hydrodynamical quantities from the simulations and showed the self-consistent emergence of expected scaling relations and spatial distribution. Here we discuss some interesting aspects regarding our results.

The strong scatter in the $L_X - \mathrm{SFR}$ relation of HMXBs is a major concern. While the expected Poissonian noise from discrete sampling of the luminosity functions was predicted by \cite{Gilfanov+2004b}, additional uncertainties arise from our SFH proxy (see Sect. \ref{sec:SSPs}). For the birthrate of massive stars (Eq. \eqref{eq:dotN}) we assumed a linear dependence on SFR that is only true as a first-order approximation and is highly sensitive to the underlying IMF. One would need to model the complete SFH of the star-forming region to account for time dependence. Since we are not able to recover this information from SSPs {in} cosmological simulations, this first-order approximation is the best we can achieve.
For the lifetime function of stars we assumed the prescription of \cite{PM1993} (PM93),  which is also implemented in the simulations \citep{LT+2007}. The advantage of the PM93 model is its simple functional form. There are more recently developed stellar lifetimes using evolutionary tracks calculated by \textsc{Parsec} \citep{Bressan2012}. Such lifetime functions are nonetheless derived for single stars despite it being known that the evolution of massive binary systems heavily depends on orbital, stellar wind, and common envelope interactions \citep[see, e.g.,][as a review on XRB evolution]{Lewin+2006}, which might drive inconsistencies in stellar age determination. For this reason, we decided to keep the PM93 model.

One of the most important drivers of uncertainty in the $L_X - \mathrm{SFR}$ relation for HMXBs at low SFR values, is the limited mass resolution of SSPs in the simulation. If the mass resolution decreases the probability of having SSPs in the required age range of $\tau_{\text{SSP}}\in[0,30]\,\text{Myr}$ becomes smaller. Because there will be fewer SSPs per galaxy to resolve SFR, it adds another layer of Poisson noise. With smaller SSP masses we would be able to more accurately estimate the HMXB population size for each galaxy as is already seen in the LMXB case where the scatter in the $L_X - M_*$ relation is much smaller. In fact, \cite{Kouroumpatzakis+2020} showed that the $L_X-\mathrm{SFR}$ relation on sub-galactic scales should be consistent with galaxy-wide scaling relations. They argue that their shallower slopes in the sub-galactic relation is caused by LMXB contamination and differences in local SFH. Nonetheless, the overall trend in our recovered $L_X-\mathrm{SFR}$ relation is still consistent with observations.

In our investigation of the metallicity dependence we concluded that there is only a weak link between the SFR and the mass-weighted stellar metallicity in the simulation since the HMXB model from \cite{Mineo+2012a} shows no significant increase in total luminosity with decreasing oxygen fraction. In fact, it instead suggests that there is a positive correlation between oxygen content and total luminosity, as indicated in Fig. \ref{fig:lx_Z_hzb}, in contrast to observations by \cite{Lehmer+2021}. We attribute this behavior of the M12 model to our small low Z sample and to the fact that our approach does not account explicitly for metallicity. Sampling from the metallicity-dependent L21 model performs as expected and correctly reproduces observational results. We note that our low SFR sample (\textit{yellow}) for $12+\log[\mathrm{O/H}] \leq 8.5$ has much higher luminosities than predicted by the Monte Carlo Markov chains of L21 for $\mathrm{SFR}\sim0.1$. Apart from the number normalization issues discussed above for low resolution SSPs, we speculate that this difference might be caused by either a different zero-point in metallicity or differences between mass-weighted stellar metallicities and gas-phase metallicities, the latter of which is used in observations. We have to compare to the mass-weighted stellar metallicity in the simulations since we rely on the metallicity estimates of the SSPs. Considering the formation process of SSPs in the \citet{SH2003} prescription and the \citet{LT+2007} metal distribution, both gas-phase and stellar metallicities should be comparable in the simulations \citep[see, e.g.,][]{Dolag+2017}.

In Fig. \ref{fig:lx_Lehm16} we compare the $L_X-\mathrm{SFR}-M_*$ relation of our sample with recent observations of galaxies within the Virgo cluster from \citet{Soria+2022} (S22).
We find excellent agreement with our data. Due to the break in our $L_X - \mathrm{SFR}$ relation for HMXBs our expected transition of LMXB to HMXB dominated galaxies, around $\log \mathrm{sSFR} \eqsim -10.5$), allows for slightly lower $L_X$ compared to a linear scaling in $L_X-\mathrm{SFR}$.
The normalization of our $L_X-\mathrm{SFR}$ relation in the limit of high sSFR appears to be higher compared to the selected studies.
In fact our model does not rule out the purely linear $L_X-\mathrm{SFR}-M_*$ relation and is equally consistent to the model in Eq. \eqref{eq:Lehm16} if we consider a lower normalization for the $L_X-\mathrm{SFR}$ relation. Interestingly, model 5 of \citet{Aird+2017} shows a sublinear normalization in the SFR component that also leads to a much better representation of the S22 data compared to the L19 model. They argue that the nonlinear SFR dependence may arise from differences in the properties of the stellar population, such as age and metallicity \citep[see, e.g.,][]{Fragos+2013a}, across a given mass range of galaxies. Combining galaxies of different redshifts thus neglects changes in stellar properties contributing to differences in the total X-ray emission per unit SFR. A similar line of argument is raised by \citet{Soria+2022} who claim that their observed incompatibility with the L19 models is caused either by the galaxy cluster environment or by unaccounted properties such as metallicity. 

\subsection{Average galaxy spectra}
When we investigated the different components contributing to galactic X-ray spectra, we only used the ideal photon lists and did not perform a full mock observation using instrumental response with subsequent unfolding of spectra. While different instruments are more or less suited to detect X-ray point sources like XRBs, we did not think that a full mock observation would prove to be beneficial in order to present the XRB algorithm. Since real observations on X-ray spectra of galaxies report unfolded spectra, we skipped the instrumental response and reconstruction in favor of cleaner and better data.
Our analysis of the average galactic spectra showed that XRBs can have significant contribution for energies $E\leq2\,\mathrm{keV}$. We note, however, that our approach suffers from the same spectral simplification as discussed in \cite{Sazonov+Khabibullin2017a}. They argue that variability of XRB spectra has to be accounted for individually including intrinsic absorption and spectral hardness for each source. Otherwise, the luminosity estimates for soft sources can be significantly different. They suggest taking ISM maps of the host galaxies into account  and individually resolving spectral shapes to improve upon luminosity estimates for XRBs. Given the modular design of our XRB module for the photon simulator \phox{}, spectral models for XRBs can easily be updated to account for XRBs in different hardness states, for example using spectral characteristics from \cite{Sazonov+Khabibullin2017c}. From the same arguments, the estimated count ratios might also be slightly different if some sources had higher flux in the soft band. 

The mean X-ray spectrum of our star-forming sample is consistent with estimates from \cite{Lehmer+2022} although their derived SED is based on observations of dwarf galaxies with sub-solar metallicity as high redshift analogs to local galaxies. The good agreement with our data is caused by our hot gas luminosity estimates being in part 0.5 dex above conventional values for the ISM (see Fig. \ref{fig:lx_sfr_gas}). An increase of gas luminosity for lower-metallicity environments from L22 thus improves the visual overlap but does not correspond to a lower metallicity in our sample. Another issue is the strong suppression of metal lines in the tabulated SED in L22. We accounted for that by convolving our derived spectrum with a fairly wide Gaussian kernel, which still does not suppress some very prominent line features at high energies. We also find that the shape of our simulated SEDs are flatter at $E\sim1-2\,\mathrm{keV}$ compared to the L22 SED. Differences in spectral shape, especially in the soft energy regime, are highly dependent on the adopted absorption models as discussed for Fig. \ref{fig:avg_spec}, which leads in our case to overestimation of the XRB component compared to values from \citet{Lehmer+2016}. In the hard band the L16 estimates are consistent with our derived values. Splitting the average spectrum into different SFR bins revealed stronger contributions from the hot gas component in high SFR galaxies as well as a higher normalization in general. This is due to the influence from massive BCG-like galaxies in the sample with hot gaseous atmospheres. In those cases, the L16 values are no longer suited to compare our derived values against. 
\citet{Gilbertson+2022} show that the expected influence of the hot gas component is also a function of stellar population age, with the hot gas contribution declining to less than 5\% in older populations, which was not investigated in our presented sample. In addition to a SFR-normalized emission model that is consistent with the L22 spectrum, \citet{Gilbertson+2022} also report stellar-mass-normalized average spectra, which we did not compare against as it goes beyond the scope of this paper. We will conduct a future, more detailed investigation of the influence of the hot gas component to the average spectrum, where we we will also drill down on line properties within the ISM and circumgalactic medium.
    
\section{Summary and conclusions} \label{sec:Conclusions}

Based on SNII rate estimates made using SSP ages and masses tracked by numerical cosmological simulations, we are able to self-consistently
reconstruct a SFR proxy suitable for Monte-Carlo-like sampling of XRB luminosity functions. We have presented a possible implementation 
within the \phox{} code and verified the validity of our approach using simulated galaxies from the       Magneticum Pathfinder simulations.
We extracted galaxy-sized halos from the simulation based on a stellar mass criterion ($10^{9.5}\,\Msun\leq M_*\leq10^{12}\,\Msun$) and produced ideal
photon lists for the hot gas and the XRB component using the stochastic photon simulator \phox{}. We assumed XRB emission followed
a static PL with photon index $\Gamma = 1.7-2$, while hot gas was modeled assuming a single temperature APEC model with 
thermal line-broadening and varying metal abundance for each gas element in the simulation. In order to account for self-absorption of the ISM at the source location, we additionally required an
intrinsic absorption component for both hot gas and the XRB component with column density $N^{\mathrm{gas}}_{H} = 5\cdot10^{21} \mathrm{cm}^{-2}$ and
$N^{\mathrm{xrb}}_{H} = 2\cdot10^{21} \mathrm{cm}^{-2}$, respectively. For each halo in our sample, we constructed
cylindrical volumes around the center of mass with $R=R_{2500}$ and $h=2\,R_{2500}$ and projected the enclosed photons, accounting for
the redshift and peculiar motion of the source with respect to the line of sight. Our findings from the resulting galactic X-ray spectra can be summarized as follows:
\begin{enumerate}
    \item The global reconstructed XRB luminosity functions perfectly resemble the underlying analytic shape.
    The observed flattening of the reconstructed XLFs below $\log L_{XRB} \approx 37$ is consistent with the expected
    one-photon luminosity. The reconstructed XLF derived from the metallicity-dependent model from \cite{Lehmer+2021}
    most closely resembles the XLF shape for solar metallicity, which is almost indistinguishable from the \cite{Mineo+2012a}
    model. A more curated sample of galaxies would be needed to reliably disentangle the effects of metallicity on
    the XLF shape, which may be analyzed in a separate study.
    \item The $L_X$-$M_*$ relation for LMXBs is tightly constrained and is in excellent agreement with the \cite{Zhang+2012} relation.
    More recent measurements of this relation by \cite{Lehmer+2020} are also consistent with our data.
    \item The $L_X$-$\mathrm{SFR}$ relation for HMXBs suffers greatly from low-number statistics for $\mathrm{SFR}\lesssim 1\,\usfr$.
    Additionally, the SFR proxy used to determine the number of HMXBs per SSP has large intrinsic uncertainties stemming from the
    limited mass resolution of the simulation and assumptions for stellar lifetime and IMF. Despite these drawbacks, 
    our approach is still consistent with a BPL relation, which is expected to arise from incomplete sampling of the
    HMXB XLF dominated by Poissonian noise (\cite{Gilfanov+2004b}).
    \item The $L_X$-$\mathrm{SFR}$-$Z$ relation as proposed by \cite{Lehmer+2021} does not arise when sampling
    from the M12a relation, indicating a weak correlation between SFR and Z within the simulation. When employing
    the L22 model directly, the correlation becomes more obvious but continues to suffer from the same uncertainties
    inherent to the HMXB sampling.
    \item The $L_X/\mathrm{SFR}$-$\mathrm{sSFR}$ relation we recovered from studying the combined effect of both XRB types
    is in remarkable agreement with recent studies of field galaxies within the Virgo cluster from \cite{Soria+2022}.
    The shape of our derived relation shows more complex behavior at $\log \mathrm{sSFR}\sim -10$ compared to the
    conventional relation found in the literature \citep{Lehmer+2010, Lehmer+2016, Lehmer+2019} due to the observed break in the HMXB relation.
    
    \item As expected, we find that HMXBs spatially coincide with the star-forming regions of their host galaxies
    and that the spatial LMXB distribution follows the stellar-mass surface density.
    \item We find that approximately 40-50\% of our galaxy sample is dominated by the combined XRB emission compared
    to emission from the hot gas component in the 0.5-8 keV energy band. In the soft band only
    30-40\% of galaxies are dominated by gas emission, while more than $90\%$ are dominated by XRB emission in the hard band. The
    surprisingly low contribution of HMXBs to these percentages reflects the relatively small portion of high SFR galaxies
    within our sample.
    \item We constructed average SFR-normalized galactic X-ray spectra from our complete star-forming sample ($\mathrm{SFR}>0.4$) and measured the relative contribution of the two XRB types and the hot gas component
    toward the total spectrum. The average spectrum of the total star-forming sample is consistent with the spectral fits from 
    \cite{Lehmer+2022} in terms of magnitude and shape, assuming an average SFR of 1. Contributions from XRBs to the soft part of the spectrum are inconsistent with values from \cite{Lehmer+2016} but consistent with those from \cite{Wik+2014}, which reflects different model assumptions for absorption and XRB spectra.
    Additionally, the normalization of the spectrum only shows minimal variations across different SFR bins and is mostly influenced by the presence of massive galaxies in the sample.
\end{enumerate}
In this paper we have presented an approach to modeling galaxy X-ray spectra from cosmological simulations. In addition to the hot gas component in the simulation, we took the properties of stellar elements into account and implemented a fast and consistent algorithm to compute XRB emission from these elements while reliably separating the two different XRB types. The presented method proves its viability as
a supporting tool for observers with its highly modular design, which also allows for alterations to employed model
spectra for individual components. Future studies that build upon the suggested implementation will improve the sample
selection and focus on the influence of X-ray AGN emission on scaling relations, on the background, and on the retrieved spectra.
\begin{acknowledgements}
We thank the anonymous referee for their helpful comments and feedback which helped improving this paper.
SVZ and VB acknowledge support by the \emph{Deut\-sche For\-schungs\-ge\-mein\-schaft, DFG\/} project nr. 415510302. KD acknowledges support through the COMPLEX project from the European Research Council (ERC) under the European Union’s Horizon 2020 research and innovation program grant agreement ERC-2019-AdG 882679.
This research was supported by the Excellence Cluster ORIGINS which is funded by the Deutsche Forschungsgemeinschaft (DFG, German Research Foundation) under Germany's Excellence Strategy – EXC-2094 – 390783311.
The calculations for the hydrodynamical simulations were carried out at the Leibniz Supercomputer Center (LRZ) under the project pr83li. We are especially grateful for the support by M. Petkova through the Computational Center for Particle and Astrophysics (C2PAP)  
\end{acknowledgements}

\bibliographystyle{aa}
\bibliography{references}

\end{document}